\documentclass[aps,prab,reprint,longbibliography]{revtex4-1}

\usepackage[english]{babel}

\usepackage{amsmath,amssymb,amsfonts}
\usepackage{graphicx}
\usepackage{algorithm}

\makeatletter
\renewcommand{\ftype@algorithm}{4}
\makeatother

\usepackage{algpseudocode}
\usepackage{booktabs}
\usepackage{bm}
\usepackage{xcolor} 

\usepackage[colorlinks=true,allcolors=blue]{hyperref}

\newcommand{\x}{\mathbf{x}}
\renewcommand{\a}{\mathbf{a}}

\newcommand{\best}[1]{\scalebox{0.94}{$\bm{#1}$}}

\begin{document}

\title{Time-Varying Bayesian Optimization and Extremum Seeking
for Drift-Tube Linac Tuning at LANSCE}

\author{Christopher Leon}
\email{cleon@lanl.gov}
\author{Diego Andrés Rivera Orona}
\author{Brad Ratto}
\author{En-Chuan Huang}
\author{Alexander Scheinker}
\affiliation{Accelerator Operations and Technology Division,
Los Alamos National Laboratory,
Los Alamos, New Mexico 87545, USA}

\begin{abstract}
Bayesian optimization (BO) has emerged as a promising machine learning approach to accelerator tuning. It has the advantage of being able to  incorporate prior information, including simulation and historical data, and explicitly model measurement noise. However, standard implementations can produce large, abrupt parameter updates that are undesirable and potentially unsafe in accelerator operation. They may also spend excessive time exploring less promising regions of the search space. We address these limitations by imposing soft and hard step size limits. We evaluate these methods by adjusting the radio frequency settings of the drift tube linac (DTL) at the Los Alamos Neutron Science Center (LANSCE) to maximize the particle count at the end of the linac. We first test the methods in simulation with constant exogenous parameters, then repeat the tests with drifting exogenous parameters. Under drifting conditions, progressively reducing the BO step size limit improved convergence. We also find that an initial BO segment followed by extremum seeking (ES), an approach we call BO-ES, combines BO's use of prior information and sample efficiency with ES's effectiveness in local optimization and drift tracking. Finally, tests on the LANSCE accelerator with BO-ES demonstrate an approximately $25\%$ reduction in losses at the end of the linac across different configurations of historical and simulation data.
\end{abstract}

\maketitle

\section{Introduction}

Bayesian optimization (BO) \cite{10.1115/1.3653121, 10.1007/3-540-07165-2_55, jones1998efficient} is well suited for optimizing black-box objective functions whose analytic form is unknown. In many practical settings, convergence and overall performance can be improved by incorporating prior information, including historical data, simulation outputs, or physics-based constraints. 

Particle accelerators pose a challenging optimization problem because of their complexity, continual drift, and unknown parameters. BO has been applied to accelerator tuning \cite{duris2020bayesian,jalas2021bayesian, kirschner2022tuning, morita2023accelerator, boltz2025leveraging}; see Ref.~\cite{roussel2024bayesian} for a review. Safety is also critical: equipment damage can be costly, and downtime reduces available beam time. These concerns have motivated research into safety-constrained accelerator optimization \cite{roussel2024bayesian, williams2024experimental}.

In this work, we focus on the important problem of optimizing the drift-tube linac (DTL) at the Los Alamos Neutron Science Center (LANSCE) \cite{scheinker2020extremum}. The DTL operating point varies over time due to temperature fluctuations, magnetic hysteresis, component replacement, and gradual wear, among other effects, such that settings that are optimal at one time may be suboptimal later. These variations occur on both long timescales (e.g., across operating periods) and short timescales (e.g., within a single run). 

Applying standard (“vanilla”) Bayesian optimization to accelerators carries substantial risks, as wildly swinging accelerator parameters can cause internal systems to take longer to regain stability or even cause catastrophic failure. 
Vanilla Bayesian optimization is also for static situations. Modifications must be made to deal with dynamic situations. While there are methods to deal with these dynamic situations, some of which we examine, there is another algorithm, bounded extremum seeking (ES) \cite{scheinker2016bounded}, which is very well-suited for the task \cite{scheinker2018demonstration,scheinker2019model,
scheinker2021extremum, scheinker2026adaptive}. ES works locally and is model-independent, therefore it is robust for quickly time-varying conditions with large distribution shifts for which memorizing prior information can be misleading and can degrade performance.

Here we introduce a novel method of combining step-size-limited Bayesian optimization with extremum seeking, which we call BO-ES. Previous work has found that combining ES and deep reinforcement learning led to improvements in controlling time-varying systems \cite{saxena2026improved}. The initial BO stage incorporates simulation and historical data to reach a promising solution basin quickly. The algorithm then switches to ES, which is a good local optimizer that can handle drifting conditions. 

First, we examine the static optimization case in simulation in Section \ref{section_label:static}. Next, the dynamic case in simulation is examined in Section \ref{section_label:dynamic}, where we have drifting exogenous parameters. Then, we test online on the LANSCE accelerator in Section \ref{section_label:LANSCE}. We conclude in Section \ref{section_label:Conclusion}.

\section{Simulating Static Parameter Mismatch}\label{section_label:static}

\begin{figure*}
\centering
\includegraphics[width=0.99\linewidth]{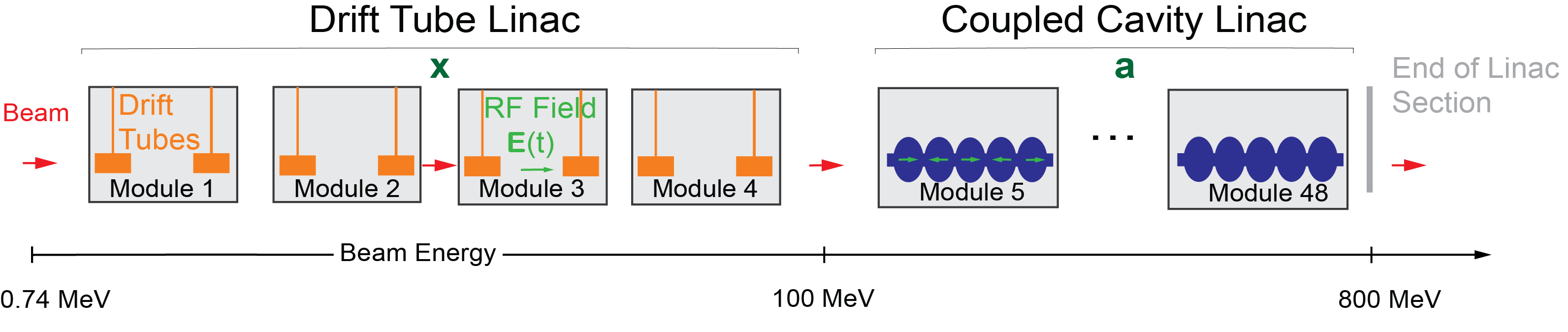}
\caption{Schematic of LANSCE DTL and CCL sections. A single DTL module consists of a large tank, which acts as a resonator. Each tank supports an oscillating electric field inside, $\mathbf{E}(t)$, with a nominal frequency of 201.25 MHz. When $\mathbf{E}(t)$ is pointing in the direction of beam propagation, synchronous particles are in the gaps between the tubes and become accelerated. When the electric field is pointing in the opposite direction, the particles are in one of several drift tubes (only two depicted), each of which acts as a Faraday cage and shields the beam from the decelerating field. The drift tubes are aligned along the tank axis and are suspended by an upper support structure. The RF settings of the DTL are denoted as $\x$. DTLs are efficient for accelerating  particles with $0.04\lesssim\beta \lesssim 0.5$ \cite{BatyginUSPAS2019_AccStruct}. At higher speeds, the CCL modules are more efficient. Each of these modules consists of multiple adjacent cavities where the oscillating electric fields (nominal frequency: 805 MHz) inside them are out of phase with their neighbors, such that synchronous particles only see an accelerating phase at each cavity (side cavities not depicted in the image). The nominal RF settings for the CCL are denoted by $\a$.  }\label{fig:schematic_DTL_CCL}
\end{figure*}

\subsection{Problem Statement}

Here we consider an optimization problem in which the accelerator model is mismatched with the physical machine. The system is assumed to operate at nominal, fixed exogenous parameters $\mathbf{a}$; however, the true exogenous parameters are $\mathbf{a}'$, which are unknown to the user. This discrepancy may arise from imperfect parameter estimation and/or from intrinsic drift in the accelerator over time. Given this model–machine mismatch, our objective is to select the control variables $\x$ to maximize the particle count measured at the end of the accelerating section.

We first evaluate the approach in simulation, where the ground-truth optimum can be reasonably estimated, enabling a quantitative performance assessment. This also allows us to identify methods that achieve high performance while maintaining conservative, safety-aware operating trajectories before deploying the approach on the physical accelerator.

While the accelerator can be treated as a complete black box, we wish to make use of the rich, physical information available from simulation data. While simulation can be incorporated via a prior mean function \cite{hwang2211prior,hwang2026customized}, we instead assume simulation is close to reality and model the residual correction. This has the advantage of clearly separating what simulation is and is not capturing. 

We model the particle-count objective, $C(\x,\a')$, using the latent surrogate:
\begin{align}\label{eqn:drift_free_model}
 \underbrace{g(\x,\a')}_{\text{full model}}
    &=
    \underbrace{f(\x,\a)}_{\text{simulation surrogate}}
    +
    \underbrace{\delta_{\a,\a'}(\x),}_{\text{sim2real correction}}
\end{align}

where:
\begin{itemize}
    \item $g(\x,\a')$ models the objective without measurement noise, with unknown $\a'$;
    \item $f(\x,\a)$ is a surrogate model for the simulation with a fixed setting of $\a$;
    \item $\delta_{\a,\a'}(\x)$ models the discrepancy between the surrogate simulation and the actual system.
\end{itemize}
Measurement noise enters through observations:
\[
    y_i=g(\x_i,\a')+\epsilon_i.
\]

We model both $f$ and $\delta_{\a , \a'}$ as Gaussian processes (GPs) \cite{rasmussen2003gaussian}, each with its own mean function and kernel (Section \ref{static_modelling}). The surrogate $f(\x , \a)$ is trained offline using simulation data, since the full high-fidelity simulator is computationally expensive and running it during operations would be infeasible. During online optimization, $f(\x , \a)$ is held frozen, while $\delta_{\a , \a'}(\x) $ gets updated with new data. 

\subsection{Data}

The synthetic data are generated using \texttt{HPSim} \cite{pang2014gpu}, with a model of LANSCE. In each accelerating cavity $i$, there is an oscillating voltage, $V_i(t)$:
$$V_i(t) = A_i \cos(2\pi f_{\mathrm{RF}} t+\phi_i),$$
where $t$ denotes physical time, $A_i$ is the voltage amplitude, $\phi_i$ is the phase, and $f_{\mathrm{RF}}=201.25$ MHz is the nominal radio frequency (RF) of the DTL. Nominal phases and amplitudes can differ from their actual machine values due to thermal fluctuations, machine wear, and other unmodeled effects.

Particles enter the DTL with $0.74$ MeV, are accelerated and exit at $100$ MeV (see Fig. \ref{fig:schematic_DTL_CCL}). While LANSCE accelerates both $H^+$ and $H^-$, we will focus exclusively on the latter, since it is the only species that makes it to the end of the linac. The control variable, $\x$, represents the RF settings of amplitude and phase for cavities of the 4 modules of the DTL: 
 $$\x = \{(A_i,\phi_i)| \ \  i=1,2,3,4\} \in \mathbb{R}^8.$$ 
RF amplitudes are expressed in the percentage units used by the LANSCE control system, and phases are expressed in degrees.

After the DTL, the $H^-$ particles enter the Coupled-Cavity Linac  (CCL) section \cite{engel1968geometrical} (see Fig. \ref{fig:schematic_DTL_CCL}). The exogenous parameters are given by the RF settings, again amplitudes and phases, for the 44 modules in the CCL:
$$\a = \{(A_i,\phi_i)| \ \ i=5,...,48\}\in \mathbb{R}^{88}.$$

In the CCL, the $H^-$ particles are accelerated to $800$ MeV. Here we train $f(\x, \a)$ on $N=2,000$ samples with a fixed $\a$ and with $\x$ being uniformly sampled between $\pm$ 3 \%/deg. of the nominal setting. The bounds were narrow as the particle count at the end of the linac section is extremely sensitive to $\x$. Many settings produced an extremely low particle count (see Appendix \ref{appendix:training_data_and_RBF}), which gives the model information about settings to avoid.

During the optimization process, the true exogenous parameters will be given by a random perturbation of the nominal values by
\begin{align}\label{eqn:exo_params}
\a'=\a+\mathbf{\eta},
\end{align}
where $\mathbf{\eta} \sim \mathrm{Unif}([-10,10]^{88})$ (\%/deg.). Note:  $\eta$ is fixed throughout the optimization process. Also, since the DTL is earlier in the acceleration process, the acceptable range for $\x$ is smaller than for $\a$.

For optimization, we wish to maximize the number of particles remaining at the end of the linac section. \texttt{HPSim} is a particle-in-cell code and we used $N_{\rm sim}=2^{20}=1,048,576$ macroparticles in all simulations. Particles remaining along track elements in the simulation data can be seen in Appendix \ref{appendix:training_data_and_RBF}.  $C(\x, \a)$ denotes the percentage of particles remaining at the end of the linac. For examples of count throughout the linac, see Appendix \ref{appendix:training_data_and_RBF}.

\subsection{Methods}\label{static_modelling}

\subsubsection{Training}
Both the simulator surrogate and the corrector are modeled as GPs:
\begin{align}
    f(\x,\a)  &\sim \mathcal{GP}\big(\mu_f(\x),\,k_f(\x,\x')\big), \\
    \delta_{\a , \a'}(\x) &\sim \mathcal{GP}\big(\mu_\delta(\x),\,k_\delta(\x,\x')\big).    
\end{align}
For both $f$ and $\delta_{\a , \a'}$ , we use the same form of the prior mean function and kernel. For the prior mean function, a constant function of $m(\x) = c$ is used, where $c$ is a hyperparameter.  We expect $f(\x , \a)$  to be a good approximation (i.e., $g(\x , \a')\approx f(\x , \a) $), and thus the correction to be small: $|f(\x , \a)| \gg |\delta_{\a , \a'}(\x) |$. Thus,  we initialize with $c_\delta =0$. However, if the data do not support this assumption, it is updated via Bayes' theorem. 

To mimic real-world noisy measurements, the online observations are corrupted by independent Gaussian noise:
\[
    y_i=g(\x_i,\a')+\epsilon_i,
    \qquad \epsilon_i\sim\mathcal{N}(0,\sigma_n^2).
\]
We add measurement noise with standard deviation $\sigma_n=2$\%. This value is not provided to the optimizer; the noise variance is inferred from the data when fitting the correction GP. Observation noise is represented separately from the latent-function kernel, through the term $\sigma_n^2 I$ in the GP likelihood and posterior equations. For the latent-function kernel, we use the RBF-$d$ ($d$-dimensional radial basis function):
\begin{align}\label{eqn:rbf-kernel}
k(\x,\x') \;=\; \sigma_f^2 \exp\!\Big(- \frac{1}{2} (\x-\x')^T \Lambda^{-1}(\x-\x')\Big),
\end{align}
where $\Lambda^{-1}=\mathrm{diag (}1/\ell^2_1,..., 1/\ell_d^2)$ and here $d=8$. This provides each dimension its own length scale, $\ell_i$. Alternatively, one can use a single length scale for all control variable dimensions. This has the disadvantage of not taking into account the different roles each $x_i$ plays. In addition, early testing showed this strategy to have less stable behavior (see Appendix \ref{appendix:training_data_and_RBF}).  Note that the simulation has some stochasticity due to numerical errors; thus, even for $f$ we have $\sigma_n^2\neq0$. However, this is generally smaller than the measurement error we will consider.

For a set of data, $ \mathcal{D} = \{ (\x_j , y_j) |j=1,..,n\} $ and input matrix $X= \{\x_j  |j=1,..,n\}$, the posterior GPs at an inquiry point  $\x_*$ have mean and covariance \cite{rasmussen2003gaussian}:
\begin{align}
\mu(\x_*) \;&=\; m(\x_*) \;+\; \mathbf{k}_{*}^T\,\big(K+\sigma_n^2 I\big)^{-1}\!\big(\mathbf{y}-\mathbf{m}_X\big),\\
\Sigma(\x_*) \;&=\; k(\x_*,\x_*) \;-\;\mathbf{k}_{*}^T\,\big(K+\sigma_n^2 I\big)^{-1} \mathbf{k}_{*},
\end{align}
where:
\begin{align}
\mathbf{y}&=\big(y_1,...,y_n\big)\nonumber\\
\mathbf{m}_X &=\big(m(\x_1),...,m(\x_n)\big)\nonumber\\
\mathbf{k}_{*}&= \big(k(\x_*,\x_1),...,k(\x_*,\x_n)\big)\nonumber\\
K&=
\begin{pmatrix}
k(\mathbf{x}_1,\mathbf{x}_1) & \cdots &k(\mathbf{x}_1,\mathbf{x}_n)\\
\vdots & \ddots & \vdots\\
k(\mathbf{x}_n,\mathbf{x}_1) & \cdots & k(\mathbf{x}_n,\mathbf{x}_n)
\end{pmatrix}.\nonumber
\end{align}
The hyperparameters of the mean function and kernel are fitted to minimize the negative log likelihood, $\mathcal{L}$:
\begin{align}\label{eqn:neg_log_lik}
    \mathcal{L}
    &=-\log p\!\left(
        \mathbf{y}\mid X,c,\boldsymbol{\ell},\sigma_f^2,\sigma_n^2
    \right) \nonumber\\
    &=\tfrac12(\mathbf{y}-\mathbf{m}_X)^\top
      (K+\sigma_n^2 I)^{-1}
      (\mathbf{y}-\mathbf{m}_X) \nonumber\\
    &\quad+\tfrac12\log\!\left[\det(K+\sigma_n^2 I)\right]
      +\tfrac{n}{2}\log(2\pi).
\end{align}
For $f$, this is done offline with the full training data. For $\delta_{\a,\a'}$, this is done every fifth iteration during online optimization, using residual targets $r_i=y_i-\mu_f(\x_i)$, where $\mu_f(\x_i)$ is the posterior mean of the frozen simulation GP at the fixed nominal exogenous parameters $\a$. A fully Bayesian method would treat the hyperparameters as distributions and integrate over them. This, however, would be prohibitively computationally expensive.

\subsubsection{Acquisition Functions}

Bayesian optimization admits a variety of acquisition functions to explore parameter space. One can sample a function from a GP posterior, analogous to sampling a vector from a multivariate normal distribution. This is done in Thompson sampling \cite{thompson1933likelihood, thompson1935theory}, which selects candidate points by drawing a sample function from the current Gaussian-process posterior: 
\begin{equation}
    \alpha^{\mathrm{TS}} \sim \mathcal{G}\mathcal{P}.
\end{equation}
Then, at time iteration $t$, the next inquiry point, $\x_t$, is the maximizer of that sample function:
\begin{align}
    \x_t = \arg \max_\x \alpha^{\mathrm{TS}}(\x).
\end{align}
During online optimization, the simulation GP remains fixed, while the correction GP is updated using residuals $r_i=y_i-\mu_f(\x_i)$. For Thompson sampling, we approximate the latent surrogate $g$ by treating the two GPs as independent and summing their posterior means and covariances. This retains the simulation surrogate's uncertainty but neglects the posterior dependence between the simulation and correction processes induced by the online observations. Measurement noise enters the correction-GP likelihood and is not added to the latent covariance used for Thompson sampling.

Relative to alternatives such as upper confidence bound (UCB), Thompson sampling is attractive because it avoids introducing an explicit exploration–exploitation hyperparameter that needs to be properly tuned (e.g., the UCB  $\beta$ parameter),  while still enjoying strong theoretical guarantees, including asymptotic optimality under certain conditions \cite{kandasamy2018parallelised}.

We consider two ways to modify the acquisition function to ensure a new inquiry, $\x_t$, does not stray too far from the previous one, $\x_{t-1}$. First, we do it with the soft constraint of Proximal Biasing \cite{roussel2021proximal}, a method we call BO-ProxBias, which modifies the acquisition function with a Gaussian factor:

\begin{equation}\label{eqn:proxbias_acq_fnc}
    \alpha^{\mathrm{PB}}(\x_{t-1},\x_{t}) = \exp \left(- \frac{||\x_{t-1}-\x_{t}||^2}{2\sigma^2 _{\mathrm{PB}}} \right) \  \times \  \alpha^{\mathrm{TS}}(\x_t).
\end{equation}
This biases the acquisitions toward nearby points. Since we need small step sizes, $\sigma_{\mathrm{PB}}$ will generally be much smaller than the scale of the global bounds. This means the acquisition values for most global points become negligibly small for Eq. \ref{eqn:proxbias_acq_fnc}. The standard practice to identify the maximizer of the Thompson sampling acquisition function is to use Sobol sequence sampling \cite{sobol1967distribution};  however, in high-dimensional settings, global Sobol sampling becomes increasingly sparse and the acquisition function is nearly all zero for all samples. This causes the maximization procedure to resemble selection from an approximately uniform distribution across the global space, leading to large jumps, which we are trying to avoid.  Substantially increasing the number of Sobol samples would mitigate this sparsity but at prohibitive computational cost. Furthermore, because the proximal factor decays as $\exp \left(- \frac{||\x_{t-1}-\x_{t}||^2}{2\sigma^2 _{\mathrm{PB}}} \right)$, candidates farther than $5 \sigma_{\mathrm{PB}}$ are down-weighted by a factor of $\sim 10^{-6}$. Given the observed scale of Thompson samples, such points are extremely unlikely to maximize the weighted acquisition. Consequently, in running BO‑ProxBias we restrict candidate generation to a local Sobol hypercube of width ±5$\sigma_{\mathrm{PB}}$ centered at the previous evaluation point, after which the Gaussian proximity weighting is applied. For this work, we use $\sigma_{\mathrm{PB}}$ = 0.1 (\%/deg.).

Alternatively, we enforce a hard per-dimension step-size limit:
\begin{equation}\label{eqn:steplmt_acq_fnc}
    \alpha^{\mathrm{SL}}(\x_{t-1},\x_{t})  = \begin{cases} \alpha^{\mathrm{TS}}(\x_{t}) & \text{if $|x_{t-1}^i-x_{t}^i| < s^i_{\rm{max}}$},  \forall i
    \\ -\infty  & \text{otherwise}. \end{cases}
\end{equation}
All step constraints are applied in raw units (percent for the amplitudes, degrees for the phases). For all $i$, we set $s^i_{\rm{max}}$ = 0.2 (\%/deg.). As with Proximal Biasing, we generate candidates only within a local hypercube centered at $\x_{t-1}$.

One could also put explicit safety limits, such as requiring the count to always remain above a certain value. In the results, we will see that having constraints on the step sizes effectively maintains a high count in our tests.

\subsubsection{Models}

\begin{algorithm}[t]
\linespread{1}\selectfont 
\caption{Step-Constrained BO with Thompson Sampling}
\label{algo:static_BO}
\begin{algorithmic}[1]
\Require Simulation data $\mathcal{D}_{\mathrm{sim}}$, parameter bounds, iterations $T$
\Require Step control $\in\{\textsc{ProxBias},\textsc{StepLmt}\}$,
         scales $\sigma_{\mathrm{PB}},s_{\max}$, refit period $K$

\State Fit simulation GP $f$ on $\mathcal{D}_{\mathrm{sim}}$ and freeze it
\State Initialize $\delta\sim\mathcal{GP}(0,k_\delta)$,
       $\mathcal{D}_\delta\gets\emptyset$,
       and $\x_0\gets\x_{\mathrm{nom}}$
       \Comment{Previous best setting}

\For{$t=1,\dots,T$}
  \State Compute the correction-GP posterior from $\mathcal{D}_\delta$
  \State Combine its mean and covariance with those of frozen $f$
         \Comment{Independent-GP approximation}
  \State Draw a Thompson sample from the combined posterior
  \If{step control $=\textsc{ProxBias}$}
    \State Select $\x_t$ using proximal bias with scale $\sigma_{\mathrm{PB}}$
           \Comment{Eq.~\ref{eqn:proxbias_acq_fnc}}
  \Else
    \State Select $\x_t$ using hard step limit $s_{\max}$
           \Comment{Eq.~\ref{eqn:steplmt_acq_fnc}}
  \EndIf
  \State Enforce parameter bounds and evaluate $y_t$ at $\x_t$
  \State $r_t\gets y_t-\mu_f(\x_t)$
  \State Append $(\x_t,r_t)$ to $\mathcal{D}_\delta$
  \State Refit correction-GP hyperparameters every $K$ iterations
\EndFor

\State $t_{\mathrm{best}}\gets\arg\max_{t\in\{1,\dots,T\}}y_t$
\State \Return $\x_{t_{\mathrm{best}}}$
\end{algorithmic}
\end{algorithm}


\begin{figure*}
\centering
\includegraphics[width=0.99\linewidth]{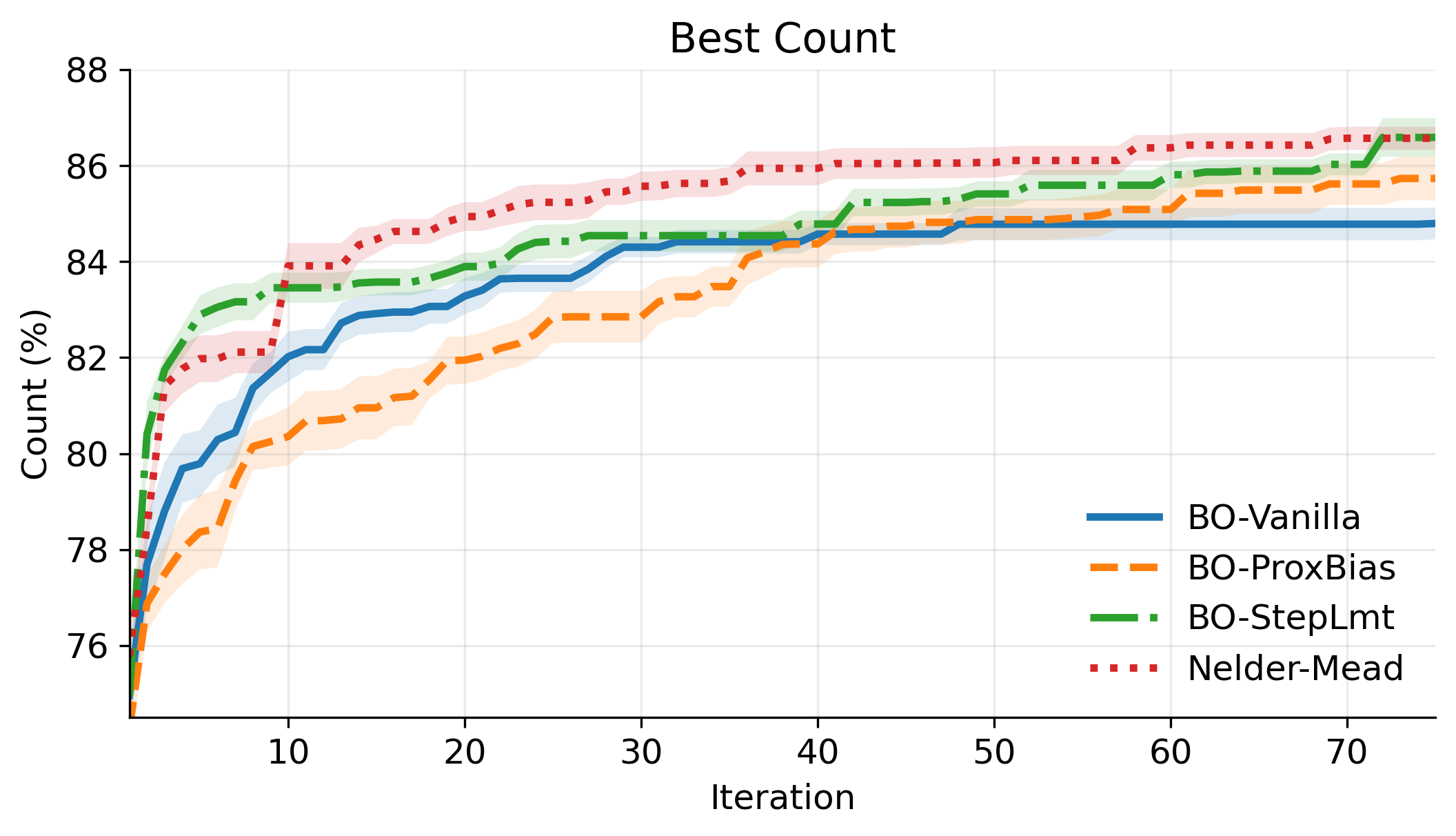}
\caption{Average of noisy count measurements using the three BO methods and Nelder-Mead. Each curve shows the mean of the 10 runs and the area shows plus/minus the standard error (${\rm SE} = \sigma/\sqrt{n}$).}\label{fig:no_drift_count}
\end{figure*}

Four methods are evaluated:
\begin{enumerate}
\item \textbf{BO-Vanilla}: standard Bayesian optimization with Thompson Sampling,
\item \textbf{BO-ProxBias}: Bayesian optimization with proximal biasing, implemented by multiplying the Thompson sampled acquisition function by a Gaussian weighting factor as in Eq. \ref{eqn:proxbias_acq_fnc},
\item \textbf{BO-StepLmt}: Bayesian optimization with explicit step-size limits on successive control updates, as in Eq. \ref{eqn:steplmt_acq_fnc},
\item \textbf{Nelder--Mead}: a classical, derivative-free optimizer that uses simplexes, included as a baseline.
\end{enumerate}


All the steps for BO-ProxBias and BO-StepLmt can be seen in Algorithm \ref{algo:static_BO}.

For the GPs and BO, all training and online optimization were done with the packages \texttt{BoTorch} \cite{botorch} and \texttt{GPyTorch} \cite{gardner2018gpytorch}. For Nelder-Mead, \texttt{scipy.optimize} \cite{2020SciPy-NMeth} was used.

\subsection{Experimental Setup}

The simulation starts by evaluating the particle count using the previous best setting, $\x_0$. Note: due to stochasticity from the numerics, each run will give different values for this. From there, each different method will run for 75 iterations (including $\x_0^*$), with ten runs each. 

\subsection{Results and Analysis}

\begin{figure*}
\centering
\includegraphics[width=0.99\linewidth]{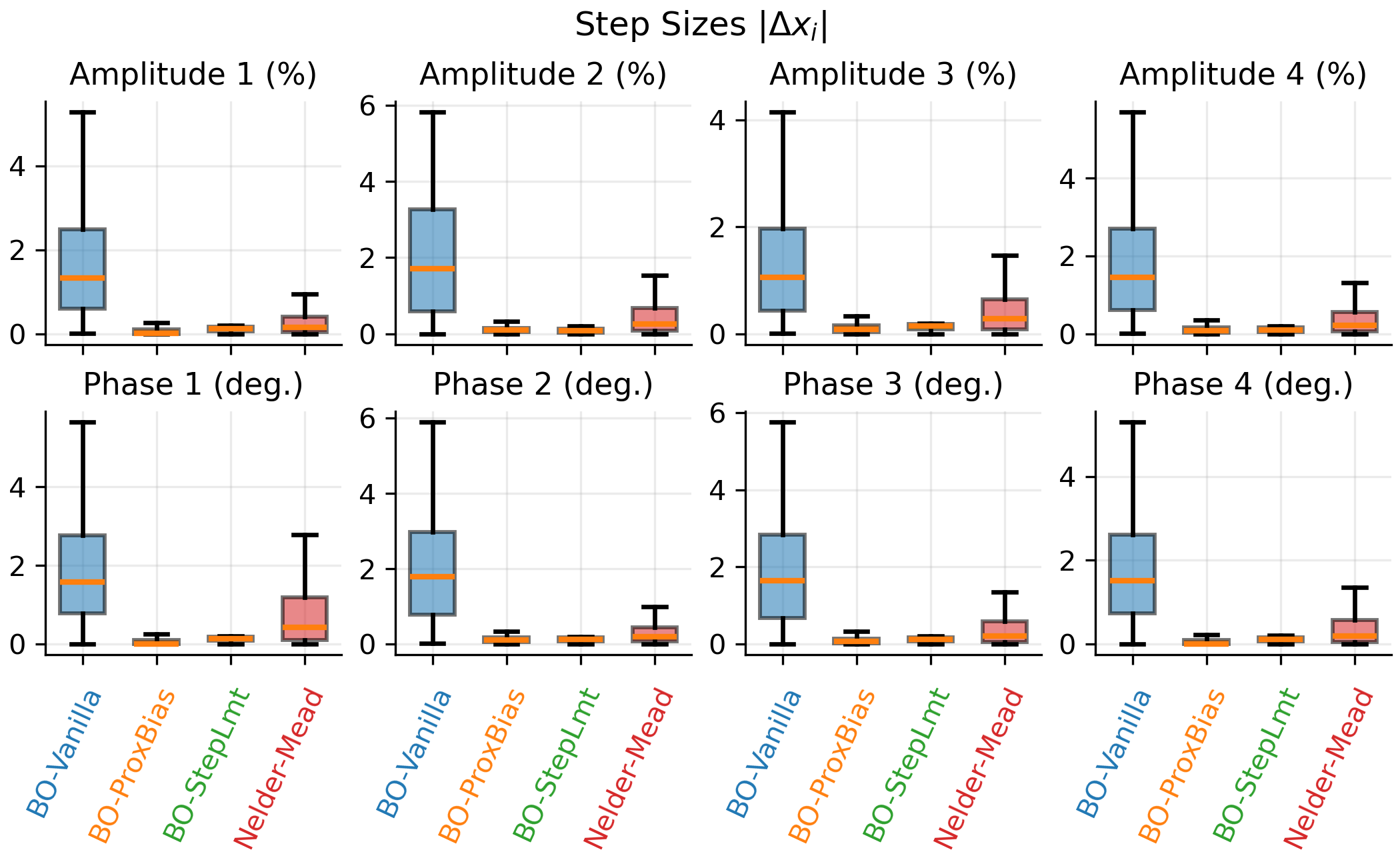}
\caption{Box plots of absolute step sizes for each method and control variable. Orange lines indicate medians; boxes span the 25th to 75th percentiles. Whiskers extend to the most extreme data points within 1.5 times the interquartile range of the box edges. }\label{fig:no_drift_step_sizes}
\end{figure*}

Figure \ref{fig:no_drift_count} shows the progression of each method. At first glance, Nelder–Mead appears to perform best, finding an optimal solution quicker. However, closer inspection reveals that this interpretation is misleading. First, as mentioned, we wish the control variables to vary slowly. Figure \ref{fig:no_drift_step_sizes} shows boxplots of the step sizes. 
Nelder-Mead generally has substantially higher step sizes than BO-ProxBias and BO-StepLmt. BO-Vanilla does even worse. 

\begin{figure*}
\centering
\includegraphics[width=0.99\linewidth]{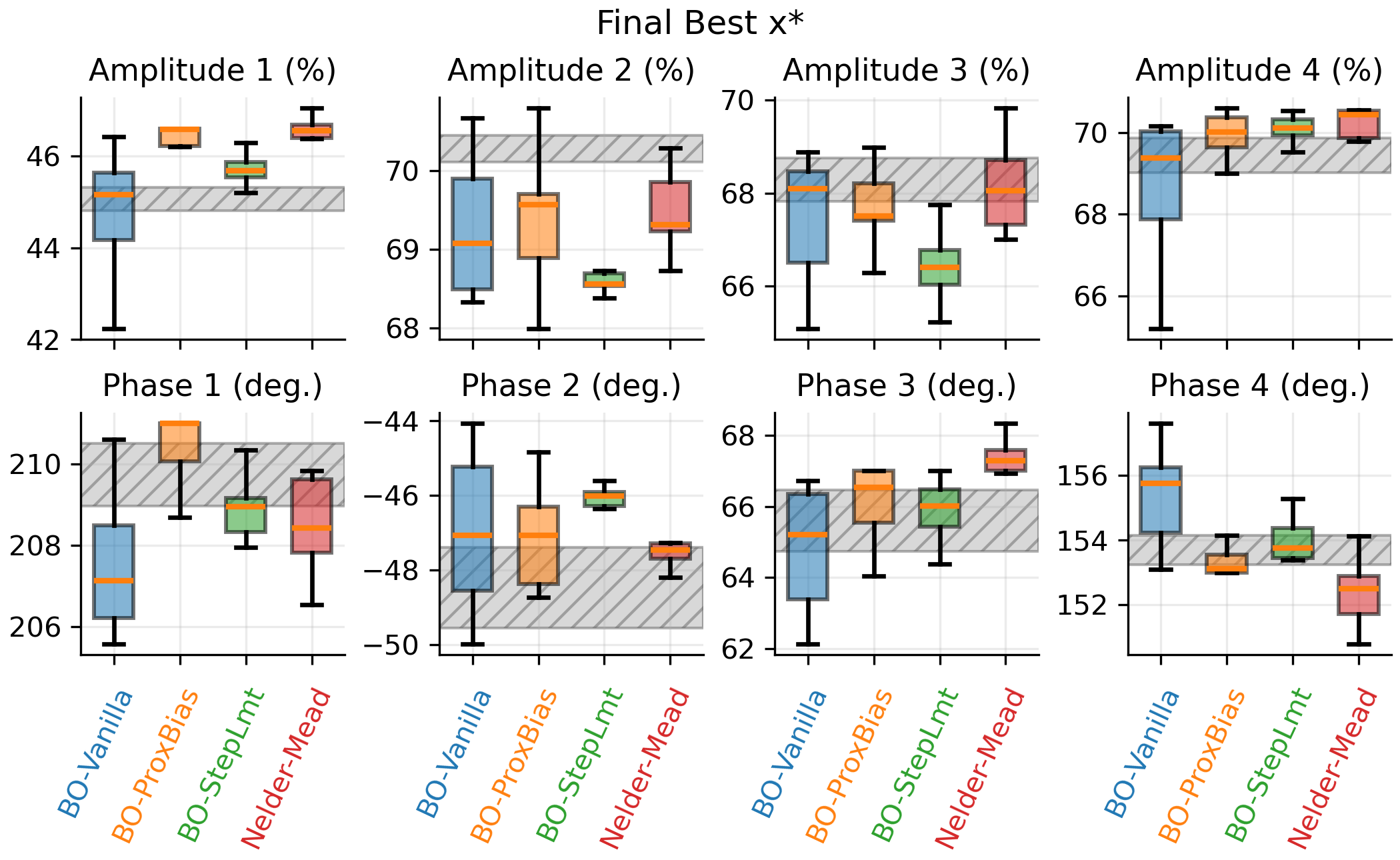}
\caption{Box plots of the 8 different variables being adjusted. The lined gray area represents the interquartile range of solutions found in the absence of measurement noise.}\label{fig:no_drift_best_x}
\end{figure*}

We can compare the solutions found with measurement noise and those found without.  We estimated a reference optimum by running BO-Vanilla with measurement noise disabled for 250 iterations over 20 trials. BO-Vanilla was used as opposed to the other methods, as we wanted to avoid the possibility of finding just a local optimum. \texttt{HPSim} is a particle-in-cell code with some stochasticity entering in from sampling of the initial distribution given the Twiss parameters and emittance and also due to the finite number of macroparticles. As such, even with measurement noise turned off, the algorithm does not find a unique optimum. Figure \ref{fig:no_drift_best_x} compares the solutions from each method against the noiseless reference distribution. Note: the simulation noise is of order $\sigma_{\rm sim.}\sim0.1-0.5 \%$, consistent with a standard deviation of $\sim 1/\sqrt{N_{\rm sim}}$. This only changes the total noise variation slightly, as given the independence of measurement and simulation noise, the BO sees a total noise variation of $\sigma_{\rm tot.} = \sqrt{\sigma^2_{\rm meas.}+\sigma^2_{\rm sim.}}\sim 2.02\%$.

Using the mean and std. of the found optima, a metric of the diagonal Mahalanobis distance squared \cite{mahalanobis2018generalized, saxena2026mahalanobis},
\begin{equation}\label{eqn:Mahalanobis_dist}
    S= \sum_{i=1}^8 \frac{(\bar{x}^*_i-\mu_{\mathrm{ref.}, i})^2}{\sigma_{ \mathrm{ref.}, i}^2},
\end{equation}
can be used to determine closeness to the solution. In Eq. \ref{eqn:Mahalanobis_dist}, $\bar{x}^*$ is the mean final best setting across trials for a method; $\mu_{\rm{ref.}}$ and $\sigma_{\rm{ref.}}$ are the per-dimension mean and standard deviation, respectively, of the reference solution distribution. Table \ref{table:Mahalanobis} shows that BO-ProxBias does best, and Nelder-Mead is much farther from the reference solution than the others. 

\begin{figure*}
\centering
\includegraphics[width=0.9\linewidth]{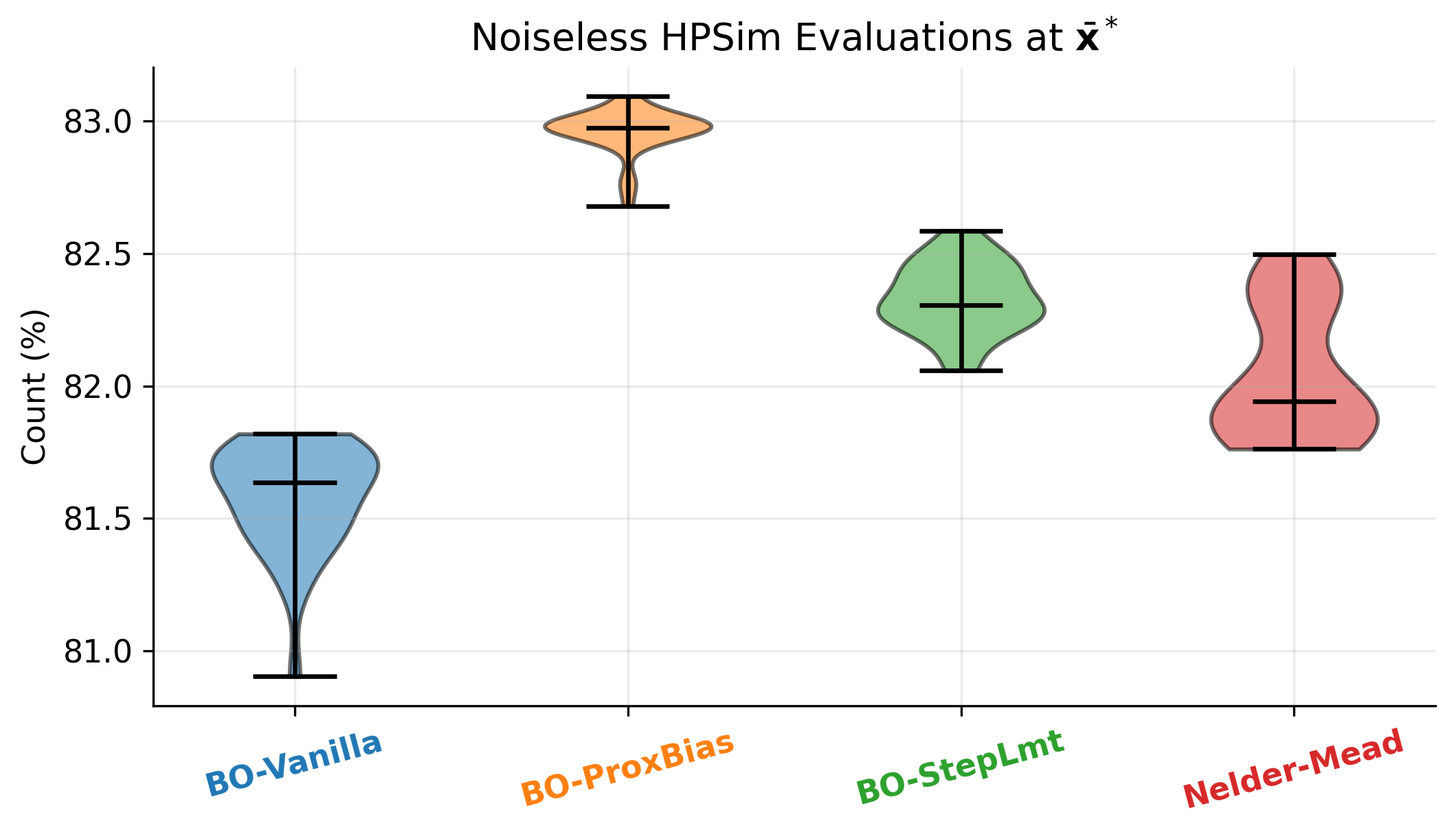}
\caption{Violin plots for the averaged best solution, $\bar{\x}^*$, for each method evaluated under no measurement noise. Black horizontal bars represent the minimum, median, and maximum of each distribution. The simulations were run 30 times for each method.  }\label{fig:no_drift_no_noise_count}
\end{figure*}

Finally, one can take the average of the best solution, $\bar{\x}^*$, found in each method and evaluate it under the condition of no measurement noise. We repeat this evaluation 10 times to account for simulation noise, and the results can be seen in Fig. \ref{fig:no_drift_no_noise_count} and Table \ref{tab:no_drift_noiseless_count}. BO-Prox does best, with BO-Step-Lmt being second, Nelder-Mead third and BO-Vanilla last. The results suggest that the high count found by Nelder-Mead was an artifact, and it was in fact chasing noise. This is borne out by the greater dispersion seen in Fig. \ref{fig:no_drift_no_noise_count}.
All BO methods take into account noise and adjust accordingly during their search. The inferior solution found by BO-Vanilla, along with its wider step sizes seen in Fig. \ref{fig:no_drift_step_sizes}, supports the criticism that standard Bayesian optimization wastes too much time exploring.


\begin{table}[t]
\centering
\linespread{1}\selectfont 
\caption{Squared diagonal Mahalanobis distance $S$
(Eq.~\ref{eqn:Mahalanobis_dist}) of $\bar{\x}^*$ for each method
from the reference obtained without measurement noise.
Boldface identifies the smallest distance. $\uparrow$ ($\downarrow$) indicates that higher (lower) values are better.}
\label{table:Mahalanobis}

\begin{tabular}{l@{\hspace{1.5em}}c}
\hline\hline
Method & $S \ \downarrow$ \\
\hline
BO-Vanilla   & $13.99$ \\
BO-ProxBias  & $\mathbf{13.08}$ \\
BO-StepLmt   & $16.65$ \\
Nelder--Mead & $26.58$ \\
\hline\hline
\end{tabular}
\end{table}

\begin{table}[t]
\centering
\linespread{1}\selectfont 
\caption{Noiseless  \texttt{HPSim} evaluation of optimized solutions.
Values are reported as mean $\pm$ standard error.
Boldface identifies the highest mean count. $\uparrow$ ($\downarrow$) indicates that higher (lower) values are better.}
\label{tab:no_drift_noiseless_count}

\begin{tabular}{l@{\hspace{1.5em}}c}
\hline\hline
Method &  \texttt{HPSim} mean count $\uparrow$ (\%) \\
\hline
BO-Vanilla   & $81.58 \pm 0.04$ \\
BO-ProxBias  & $\mathbf{82.96 \pm 0.02}$ \\
BO-StepLmt   & $82.33 \pm 0.02$ \\
Nelder--Mead & $82.05 \pm 0.05$ \\
\hline\hline
\end{tabular}
\end{table}

\section{Simulating Dynamic Parameter Mismatch}\label{section_label:dynamic}

\subsection{Problem Statement}

The problem is similar to that in the previous section, except now the exogenous parameters drift with time. Thus, in addition to the fixed perturbation $\boldsymbol{\eta}$ in Eq.~\ref{eqn:exo_params}, there is an iteration-dependent drift, $\mathbf{d}(t)$:
\begin{align}\label{eqn:exo_params_drift}
\a'(t)=\a+\mathbf{d}(t)+\boldsymbol{\eta}.
\end{align}
Here, exactly how the system is perturbed and how it is drifting are unknown to the optimizer; only the noisy measurements are known. 

\subsection{Data}
We again work in simulation.
Two drift terms that mimic real-world changing conditions will be considered:
\begin{align}
    \textbf{Linear:}& \quad d_i(t)=b_i t,\\
    \textbf{Sinusoidal:}& \quad
    d_i(t)=D_i\big(\sin(2\pi f_i t+\Phi_i)-\sin\Phi_i\big),
    \label{eqn:drift_sin}
\end{align}
with parameterization chosen such that $d_i(0)=0$. For each exogenous parameter, we assign values randomly and uniformly:
\begin{align}
    b_i &\sim \mathrm{Unif}([0.025,0.075]), \nonumber\\
    D_i &\sim \mathrm{Unif}([0.25,1.5]), \nonumber\\
    f_i &\sim \mathrm{Unif}([0.01,0.025])
    \Rightarrow T_i^{\rm drift}\in[40,100], \nonumber\\
    \Phi_i &\sim \mathrm{Unif}([0,2\pi]). \nonumber
\end{align}
Here, $b_i$ is the rate at which the $i^{\rm th}$ parameter is drifting in the linear case, $D_i$ is the sinusoidal drift amplitude, $f_i$ is its frequency in cycles per iteration, and $T_i^{\rm drift}=1/f_i$ is its period in iterations. Tuning is now more difficult than before, as the 88 exogenous variables are all changing, each with its own linear rate or amplitude, frequency, and phase.

\subsection{Methods}

Some of the previous methods were adapted for the dynamic case. Since Nelder-Mead is designed for static optimization, we do not employ it here.

\subsubsection{Time-Varying BO}

We explore  time variation by exponentially weighting the data to emphasize more recent observations. In addition, an exponentially decaying step-size limit is introduced, with a half-life of 50 iterations. This was done because BO did not settle with a constant step-size limit in our tests. One possible explanation is excessive exploration, an issue that motivated the localized search used in trust-region Bayesian optimization (TuRBO) \cite{eriksson2019scalable}. TuRBO adjusts its trust-region size adaptively, whereas our exponential decay schedule was selected empirically.

\subsubsection{Extremum Seeking}

\begin{table*}[t]
    \centering
    \linespread{1}\selectfont 

    \caption{Comparison of Bayesian optimization and bounded
    extremum seeking. Here $n$ is the number of observations
    used to fit the Gaussian process.}
    \label{tab:bo_es_comparison}

    \renewcommand{\arraystretch}{1.15}
    \setlength{\tabcolsep}{4pt}

    \begin{ruledtabular}
    \begin{tabular}{p{0.22\linewidth} p{0.35\linewidth} p{0.35\linewidth}}
        \multicolumn{1}{c}{Characteristic}
        & \multicolumn{1}{c}{Bayesian optimization}
        & \multicolumn{1}{c}{Extremum seeking (bounded)} \\
        \hline

        Search scope
        & Global
        & Local \\

        System knowledge
        & Prior information can be incorporated
        & Model-free \\

        Step-size limits
        & Can be imposed
        & Inherent step size \\

        Drift tracking
        & Requires model adaptation
        & Naturally tracks drift \\

        Computational burden
        & High; standard GP fitting scales as $\mathcal{O}(n^3)$
        & Low; simple feedback update \\

    \end{tabular}
    \end{ruledtabular}
\end{table*}
Like Bayesian optimization, extremum seeking is an optimization method used when the analytic form of a function is unknown; see Table \ref{tab:bo_es_comparison} for a comparison against BO. Coming from control theory, the approach oscillates the control variables, each with its own frequency. In the asymptotic limit of high frequencies, the average behavior of the system replicates gradient descent, despite the exact gradients not being directly available. The approach has a number of nice theoretical properties \cite{scheinker2017model} and has been demonstrated to work in accelerator applications \cite{williams2024experimental, scheinker2018demonstration, scheinker2021extremum, scheinker2021adaptive, roussel2023xopt}. 

The continuous ES dynamics are:
\begin{equation}\label{eqn:ES-varying}
    \frac{dx^i}{d\tau}
    =\sqrt{2\alpha\omega_i}
    \cos\big(\omega_i\tau+\kappa J\big),
\end{equation}
where $\tau$ is a continuous algorithmic time parameter, rather than physical time or the overall iteration index $t$. Here, $\alpha$ controls the dithering amplitude, $\omega_i$ is the angular frequency for the $i^{\rm th}$ parameter, $\kappa>0$ is the ES gain, and $J$ is the objective to be minimized. We let $\omega_i=r_i\omega$ for some $\omega$ and require that $r_i$ and $r_j$ not be integer multiples for $i\neq j$, to avoid resonances. For particle-count maximization, we use $J=-C$.

On average, Eq.~\ref{eqn:ES-varying} produces gradient descent on $J$. For a proof sketch of why this holds in the 1D case, see Appendix \ref{appendix:proof_ES_grad_desc}. For the multidimensional case and a more rigorous treatment, see Ref.~\cite{scheinker2017model}.

For the discrete implementation, let $j=1,2,\ldots, n_{\rm ES}$.
Define:
\[
    \tau_j=j\Delta\tau,
    \qquad
    \Delta\tau=\frac{2\pi}{10\omega_{\max}},
    \qquad
    \omega_{\max}=\max_i\omega_i.
\]
We update the $i^{\text{th}}$ component according to:
\begin{equation}\label{eqn:ES-vary-discrete}
    x^{i}_j
    =x^{i}_{j-1}
    +\Delta\tau\sqrt{2\alpha\omega_i}
    \cos\big(\omega_i\tau_j+\kappa J_{j-1}\big),
\end{equation}
where $J_{j-1}$ is the measured objective at the preceding setting $\x_{j-1}$. Thus, the first ES update uses the phase $\omega_i\Delta\tau$ and the measurement at the initial ES setting. The ES updates were calculated by mapping the accelerator settings $x_j^i$ onto the interval $[-1,1]$. This was done using the average and range of the global bounds.

Defining $\Delta x^{i}_j=x^{i}_j-x^{i}_{j-1}$, Eq.~\ref{eqn:ES-vary-discrete} gives:
\begin{align}
    |\Delta x^{i}_j|
    &\leq\Delta\tau\sqrt{2\alpha\omega_i}
    \nonumber\\
    &\leq\frac{2\pi}{10}
    \sqrt{\frac{2\alpha}{\omega_{\max}}}.
    \nonumber
\end{align}
Thus, like BO-StepLmt, ES has a hard step-size limit.

The oscillation phase can be written as:
\begin{align}
    \omega_i\tau_j
    =\frac{2\pi\omega_i}{10\omega_{\max}}j.
\end{align}
Therefore, the angular frequency per ES update is
$\omega_i^{\rm iter}=2\pi\omega_i/(10\omega_{\max})$.
This determines how quickly the $i^{\text{th}}$ parameter oscillates when $J$ is constant.

\subsubsection{Hybrid: BO-ES}

\begin{algorithm}[t]
\linespread{1}\selectfont 
\caption{BO-ES}
\label{alg:bo-es}
\begin{algorithmic}[1]
\Require Frozen simulation GP $f$, parameter bounds
\Require Switch iteration $n_{\mathrm{switch}}$, total iterations $T$
\Require Initial movement scale $s_0$, halving interval $H$, refit period $K$
\Require Step control $\in\{\textsc{StepLmt},\textsc{ProxBias}\}$; ES parameters

\State Initialize $\x_0$ at the previous best setting,
       $\delta\sim\mathcal{GP}(0,k_\delta)$,
       and $\mathcal{D}_\delta\gets\emptyset$

\For{$t=1,\dots,T$}
  \If{$t\le n_{\mathrm{switch}}$} \Comment{BO phase}
    \State $s_t\gets s_0\,2^{-t/H}$ \Comment{Exponential decay}
    \State Compute the correction-GP posterior using age-weighted data
    \State Combine its mean and covariance with those of frozen $f$
           \Comment{Independent-GP approximation}
    \State Draw a Thompson sample from the combined posterior
    \State Select $\x_t$ using the chosen step control at scale $s_t$
           \Comment{Hard limit or proximal bias}
  \Else \Comment{ES phase}
    \State $j\gets t-n_{\mathrm{switch}}$,
           \quad $J^{\rm ES}_{j-1}\gets-y_{t-1}$
    \State Take an ES step from $\x_{t-1}$ using
           Eq.~\ref{eqn:ES-vary-discrete}
           \Comment{Starts from the last BO iterate}
  \EndIf

  \State Enforce parameter bounds and applicable step limits
  \State Evaluate the system at $\x_t$ to obtain $y_t$

  \If{$t\le n_{\mathrm{switch}}$}
    \State $r_t\gets y_t-\mu_f(\x_t)$
    \State Append $(\x_t,t-1,r_t)$ to $\mathcal{D}_\delta$
           \Comment{Timestamp for age weighting}
    \State Refit correction-GP hyperparameters every $K$ iterations
           using age-weighted data
  \EndIf
\EndFor
\end{algorithmic}
\end{algorithm}

BO and ES offer complementary advantages. BO can exploit prior information through its surrogate model to accelerate convergence, while ES is a model-free feedback method that is particularly well-suited to systems with time-varying or drifting optima. This motivates a hybrid strategy for our case: 

\begin{enumerate}
    \item Initiate search using BO with step-size constraints.
    \item After $n_{\mathrm{switch}}$ BO iterations, switch to ES.
\end{enumerate}
For BO-ES, the ES update number in Eq. \ref{eqn:ES-vary-discrete} is $j=t-n_{\mathrm{switch}}$. For the pseudocode implementation, see Algorithm \ref{alg:bo-es}.

In this framework, BO compensates for long-term, run-to-run drift and provides a good warm start, after which ES takes over and provides continuous within-run adaptation by persistently perturbing and updating the control parameters during operation. A further practical advantage is scalability: ES can operate indefinitely, while standard BO implementations based on Gaussian processes become increasingly computationally expensive as the number of iterations grows, due to the $\mathcal{O}(n^3)$ cost of the required kernel matrix inversion. Furthermore, in settings where step size is not an issue, BO can start with a global search and ES can then hone in on a promising location|a potential global optimum. The hybrid strategy without step limits for BO has been used before for this task in minimizing power consumption for a vapor compression system \cite{chakrabarty2021accelerating}. 

\subsubsection{Models}

The following four models are run for this setup:
\begin{enumerate}
\item \textbf{No Update}: remains at the previous optimal point, $\x_0^*$. This baseline method also shows how the system is fluctuating.
\item \textbf{ES}: Extremum seeking with $\kappa=32$, $1\leq\omega_i\leq1.75$, and $\alpha=1$.
\item \textbf{BO}: Bayesian optimization with an exponentially decreasing movement scale: $s(t)=0.1\times2^{-t/50}$, where $t$ is the overall iteration index. Also, an exponential iteration weighting is used to discount older data.
\item \textbf{BO-ES}: Run BO through iteration $n_{\mathrm{switch}}=40$, then begin ES at iteration $n_{\mathrm{switch}}+1$. Hyperparameters are the same as for ES and BO. 
\end{enumerate}

Furthermore, both BO and BO-ES will be tested with the different step size constraints of Step Limit or Proximal Bias. 
 
\subsection{Results and Analysis}

\begin{figure*}
\centering
\includegraphics[width=0.99\linewidth]{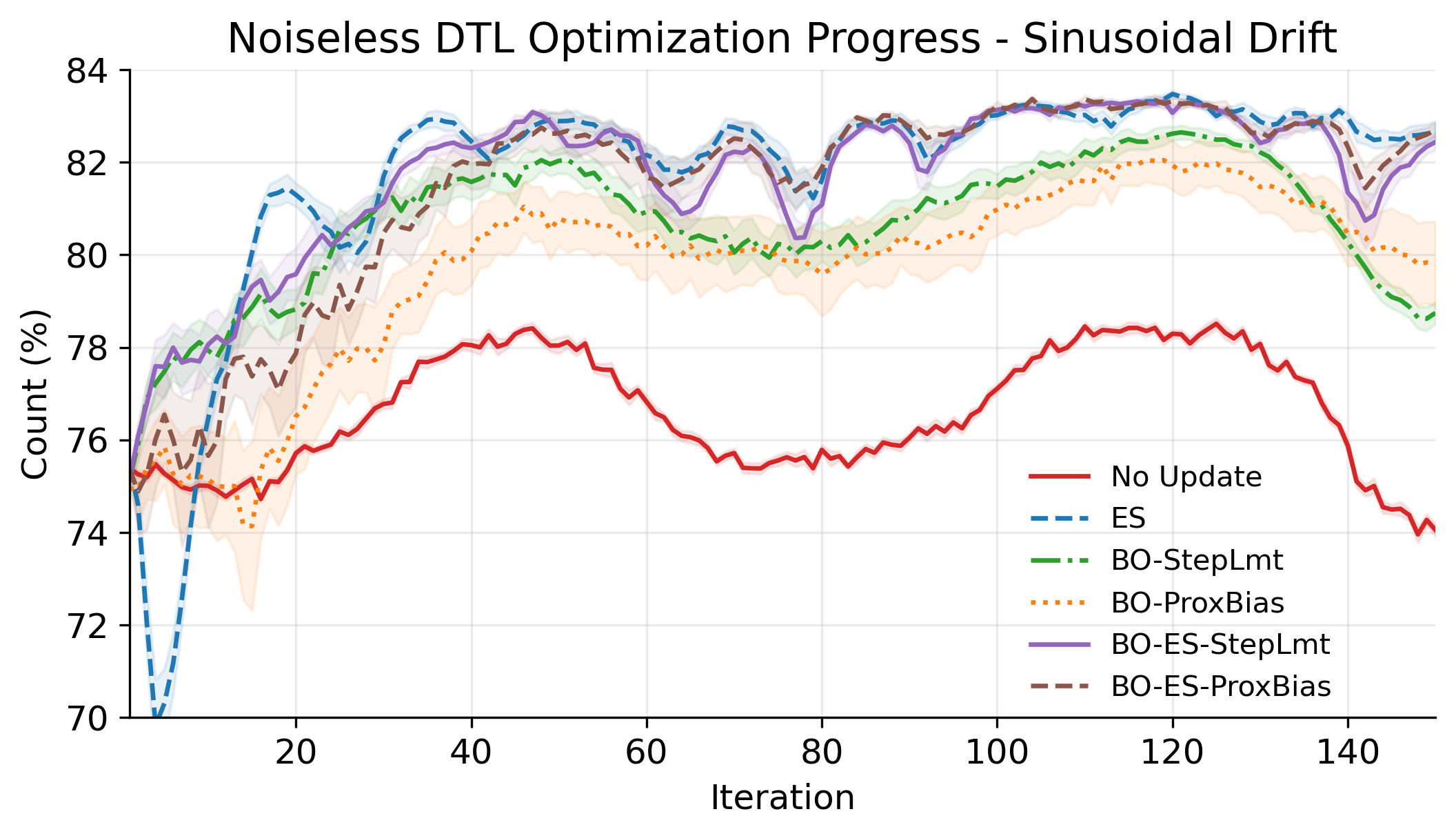}
\caption{Noiseless progress for sinusoidal drift. Displayed are the mean and standard error  (${\rm SE} = \sigma/\sqrt{n}$) of the 10 runs for each method.}\label{fig:drift_noiseless_progress}
\end{figure*}

Progress for each method in the sinusoidal case can be seen in Fig. \ref{fig:drift_noiseless_progress}. The noiseless count is shown (for a comparison of the noisy vs noiseless measurements, see Appendix \ref{appendix:noisy_measurement}). Table \ref{tab:drift_statistics_sin} shows several metrics of the optimization for each method over all 10 runs. The No Update method can be seen to be sinusoidal with a period of $T\approx 75$. A first-order expansion of $C(\x,\a')$ with perturbative terms given by Eq. \ref{eqn:drift_sin} will give a sum of sines. We can expect a frequency of $T \sim  1/\Delta f ~ \sim 67$, which is in line with the curve. All other methods do significantly better than the No Update baseline. 

ES and the BO-ES methods settle to a similar plateau after iteration $\sim$40. However, the count under ES in the beginning actually decreases and reaches a low count. This can be seen in Table \ref{tab:drift_statistics_sin}, where the Bottom 1\% of the runs for BO-ES-StepLmt is significantly higher than that of ES. ES eventually corrects itself and matches the BO-ES methods even before the switch, but that initial decrease is undesirable. Meanwhile, all methods using BO generally increase monotonically, though for each individual run this was not necessarily the case. However, the pure BO methods struggle with handling drift relative to the ES and BO-ES runs. The observed BO behavior may arise from the decreasing step size, but incorporating this feature nonetheless yielded superior performance. Thus, BO-ES makes good use of the surrogate simulation model, even though the current state of the accelerator does not match the offline training data exactly. It also deals well with drifting conditions. 

\begin{figure*}
\centering
\includegraphics[width=0.99\linewidth]{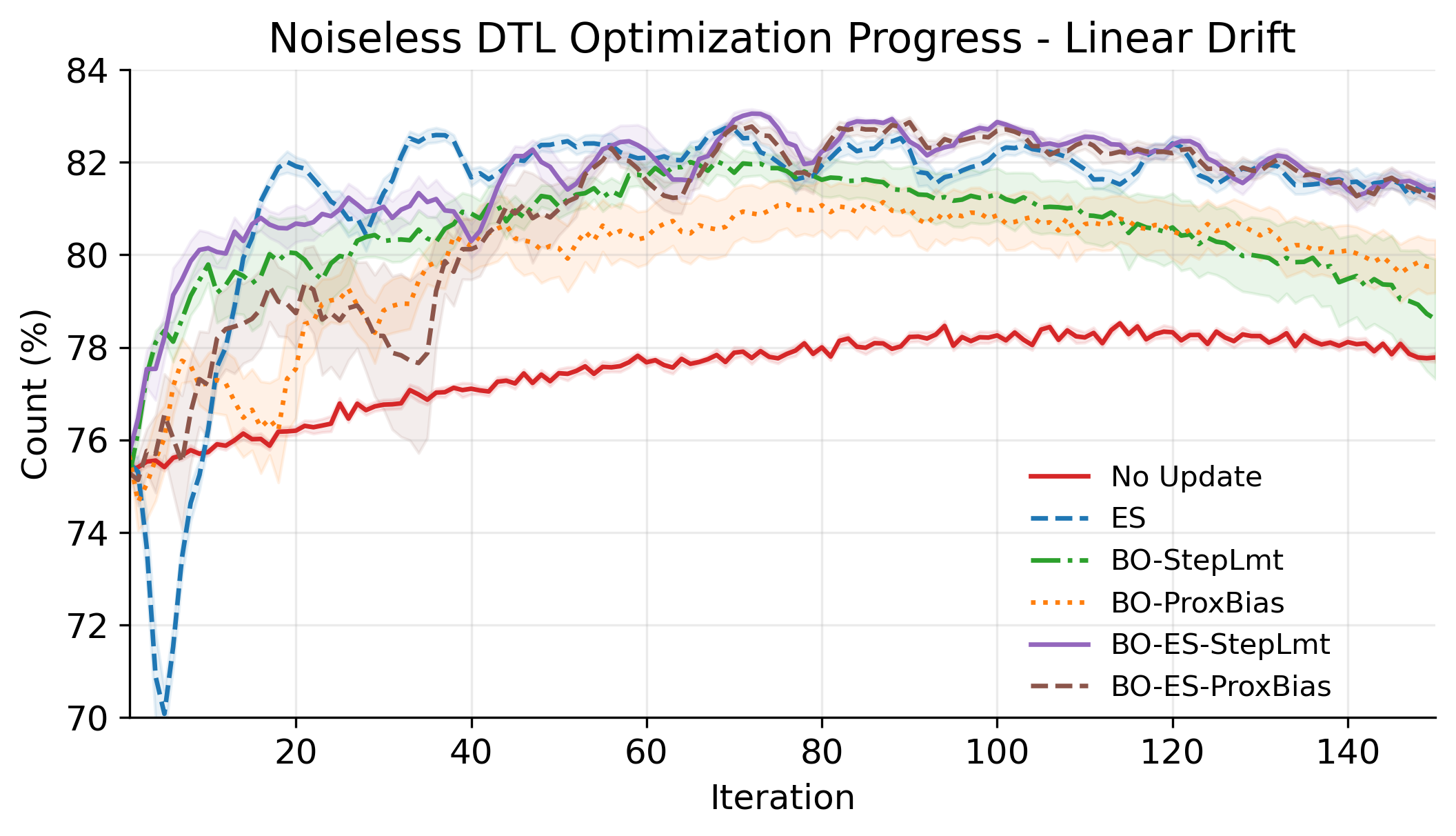}
\caption{Noiseless progress for linear drift. Displayed are the mean and standard error (${\rm SE} = \sigma/\sqrt{n}$) of the 10 runs for each method.}\label{fig:drift_noiseless_progress_lin}
\end{figure*}

The results for linear drift follow a similar pattern, as can be seen in Fig.~\ref{fig:drift_noiseless_progress_lin}. The BO methods exhibit on average a nearly monotonic increase in particle count during the initial phase. However, the pure BO methods generally have more variability, as can be seen with the wider uncertainty bands in the later stages relative to the other methods. Although they outperform the No Update baseline on average, their final particle counts are comparable to those of the baseline within the uncertainty bounds. BO-ES-StepLmt again achieves high particle counts both on average and in the bottom 1\%, as seen in Table \ref{tab:lin_drift_statistics}.



\begin{table*}[t]
\centering
\linespread{1}\selectfont 
\caption{DTL optimization statistics under sinusoidal drift.
$C_t$ is the particle count percentage at iteration $t$,
and $\Delta C_t=C_t-C_{t-1}$.
Boldface identifies the best-performing methods, including
multiple methods in cases of a statistical tie. $\uparrow$ ($\downarrow$) indicates that higher (lower) values are better.}
\label{tab:drift_statistics_sin}

\setlength{\tabcolsep}{5pt}
\begin{tabular}{lccccc}
\hline\hline
Method
& \shortstack{Avg. $C_t$ $\uparrow$\\ (all, \%)}
& \shortstack{Avg. $C_t$ $\uparrow$\\ (after 50, \%)}
& \shortstack{Avg. $|\Delta C_t|$ $\downarrow$ \\ (all, \%)}
& \shortstack{Avg. $|\Delta C_t|$ $\downarrow$ \\ (after 50, \%)}
& \shortstack{Bottom 1\%  $\uparrow$ \\  (\%)} \\
\hline
No Update
& $76.63 \pm 0.03$ & $76.73 \pm 0.04$
& $0.47 \pm 0.01$ & $0.48 \pm 0.01$ & $73.88$ \\
ES
& \best{81.73 \pm 0.07} & \best{82.70 \pm 0.02}
& $0.42 \pm 0.02$ & \best{0.26 \pm 0.02} & $69.83$ \\
BO-StepLmt
& $80.68 \pm 0.05$ & $81.11 \pm 0.05$
& \best{0.38 \pm 0.01} & \best{0.29 \pm 0.01} & \best{75.11} \\
BO-ProxBias
& $79.71 \pm 0.08$ & $80.73 \pm 0.07$
& $0.49 \pm 0.05$ & \best{0.31 \pm 0.05} & $69.08$ \\
BO-ES-StepLmt
& \best{81.70 \pm 0.05} & $82.40 \pm 0.03$
& $0.41 \pm 0.01$ & $0.35 \pm 0.02$ & \best{75.11} \\
BO-ES-ProxBias
& $81.47 \pm 0.07$ & $82.60 \pm 0.03$
& $0.48 \pm 0.04$ & \best{0.27 \pm 0.02} & $70.84$ \\
\hline\hline
\end{tabular}
\end{table*}

\begin{table*}[t]
\centering
\linespread{1}\selectfont 
\caption{DTL optimization statistics under linear drift.
$C_t$ is the particle count percentage at iteration $t$,
and $\Delta C_t=C_t-C_{t-1}$.}
\label{tab:lin_drift_statistics}

\setlength{\tabcolsep}{5pt}
\begin{tabular}{lccccc}
\hline\hline
Method
& \shortstack{Avg. $C_t$ $\uparrow$\\ (all, \%)}
& \shortstack{Avg. $C_t$ $\uparrow$\\ (after 50, \%)}
& \shortstack{Avg. $|\Delta C_t| $ $\downarrow$\\ (all, \%)}
& \shortstack{Avg. $|\Delta C_t|$ $\downarrow$\\ (after 50, \%)}
& \shortstack{Bottom 1\% $\uparrow$ \\ (\%)} \\
\hline
No Update
& $77.51 \pm 0.02$ & $78.02 \pm 0.01$
& $0.40 \pm 0.01$ & $0.41 \pm 0.01$ & $75.63$ \\
ES
& $81.29 \pm 0.06$ & $81.99 \pm 0.02$
& $0.40 \pm 0.02$ & \best{0.25 \pm 0.02} & $75.70$ \\
BO-StepLmt
& $80.53 \pm 0.06$ & $80.88 \pm 0.07$
& \best{0.37 \pm 0.02} & \best{0.27 \pm 0.02} & $75.78$ \\
BO-ProxBias
& $79.84 \pm 0.06$ & $80.56 \pm 0.06$
& $0.46 \pm 0.04$ & \best{0.29 \pm 0.05} & $75.88$ \\
BO-ES-StepLmt
& \best{81.64 \pm 0.04} & \best{82.21 \pm 0.03}
& \best{0.35 \pm 0.02} & \best{0.25 \pm 0.02} & \best{78.74} \\
BO-ES-ProxBias
& $80.93 \pm 0.07$ & $82.07 \pm 0.03$
& $0.42 \pm 0.04$ & \best{0.25 \pm 0.02} & $75.62$ \\
\hline\hline
\end{tabular}
\end{table*}

\section{Optimizing the DTL at LANSCE}\label{section_label:LANSCE}

\subsection{Problem Statement}

We selected BO-ES-StepLmt for testing on the LANSCE accelerator based on its effective performance in simulation and its guaranteed hard step size limit. Deployment required some modifications to meet the stricter safety constraints of live accelerator operations.

First, following operational recommendations, we looked to minimize linac loss rather than to maximize particle count. This was primarily due to the loss monitors having better sensitivity than the current monitors. Second, we revised the formulation to incorporate both simulation and historical data, accounting for the latest setpoints following the accelerator's recent calibration. Third, we adjusted the step size limits to comply with existing safety guidelines: a phase step limit of $0.05\ \rm{ deg.}$ and an amplitude step limit of $0.02\%$. Global limits for the parameters were also narrower than in simulation: a global amplitude range of $0.25 \%$ and a global phase range of $1.0$ deg. This is compared to $\pm 3 \%/{\rm deg.}$ for the simulation data. Finally, following an operational recommendation to remove unnecessary degrees of freedom, we kept the phase of Tank 1 fixed. 

We investigated the timescale over which historical data were most useful by considering periods of one week, one month, and three months.

\subsection{Data}

The same simulation data as in the preceding sections were used to train the GPs. Historical data were preprocessed to retain only observations collected while the accelerator was operational, with each dataset containing 1,000 observations from the corresponding period. Operational safety requirements restricted the range of parameter values in the historical data, whereas simulations provided data over a much wider parameter range.

The historical data were archived only when the phase change was more than  $0.01\ \rm{deg.}$ or the amplitude by more than $0.1\%$, which limited the calibration to rough accuracy. 

During testing on the LANSCE accelerator, we paused for three seconds after each iteration to allow the accelerator to settle. The linac loss is obtained by summing the readings of 88 loss monitors throughout the 44 CCL modules. Limited beam development time allowed only one run per method.

\subsection{Methods}

\begin{figure*}
    \centering
    \includegraphics[width=\linewidth]{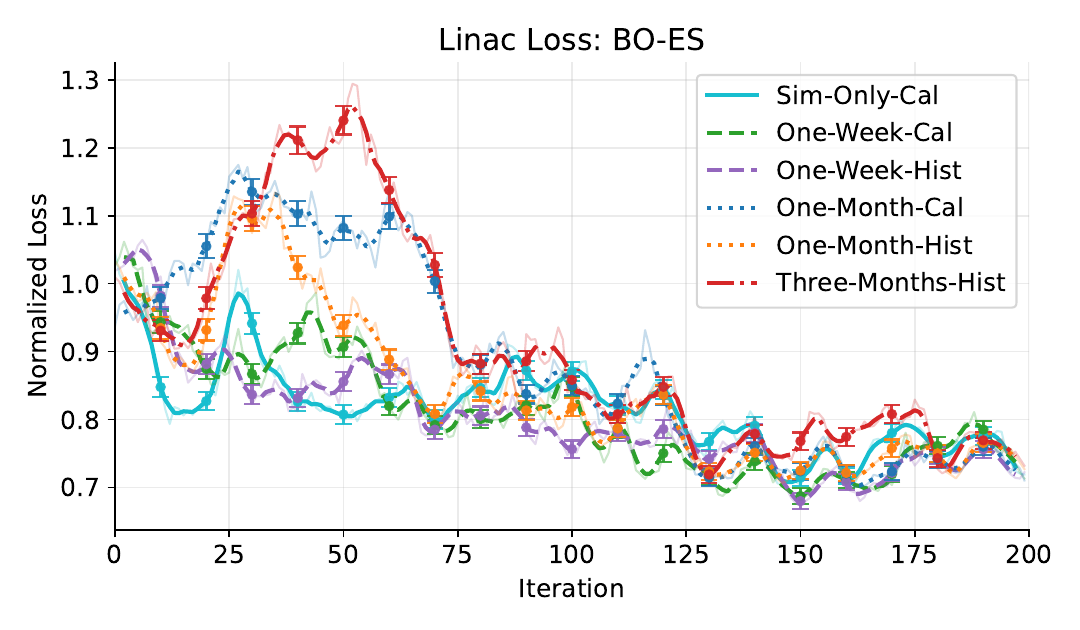}
    \caption{Five-point window average of the normalized loss (normalized by $\bar L_0$) per iteration. Faint lines in the backdrop show the raw value for each iteration. Error bars represent $1\sigma$ of uncertainty for the window average and are only displayed every $10^{\rm th}$ point for visual clarity. The dash-dotted line (red) shows measured loss using three months of historical data to train the GP for the BO. The dotted (blue and orange) lines show measured loss using one month of historical data to train the GP. In this case, we see that the calibrated GP yields considerably worse results, particularly when we look between the $25^{\rm th}$ and $75^{\rm th}$ iterations. The dashed (green and purple) lines show measured loss using one week of historical data for the GP training. We see that for these one-week GPs, the calibrated and purely historical GPs perform relatively equally over all iterations. Finally, the simulation-only GP (solid, cyan) gave the best results across the first few BO iterations, but was eventually outperformed by the one-week GPs (see Table \ref{tab:LANSCE_results} for a comparison). A good benchmark for overall loss minimization is performance up to the $50^{\rm th}$ iteration, when we change from BO to ES. We see that only the One-Month-Calibrated and Three-Months-Historical GPs have performed genuinely poorly: indeed, they increased loss up to this benchmark. See Fig. \ref{fig:cumulative_loss} for behavior past this benchmark.}
    \label{fig:Linac_Loss_BO_ES}
    \label{fig:LANSCE_loss}
\end{figure*}

First, we trained separate latent GPs, $g \sim \mathcal{GP}$, to model beam loss using only historical data spanning one week, one month, and three months.

In addition, we also trained a single GP, $f$, on simulation data to predict particle count as before but now without $\phi_1$:  $\x\in \mathbb{R}^7$. To map the predicted count percentage to loss and incorporate historical data, we fitted the following latent model:
\begin{equation}\label{eqn:sim_plus_historical}
    g(\x) = c_0 + c_1 f(\x-\mathbf{\Delta})
    + \delta^{\rm hist}(\x).
\end{equation}
The historical loss observations satisfy
\[
    L_i^{\rm hist}=g(\x_i)+\epsilon_i^{\rm hist},
    \qquad
    \epsilon_i^{\rm hist}\sim
    \mathcal{N}(0,\sigma_{\rm hist}^{2}).
\]
In Eq. \ref{eqn:sim_plus_historical}, $c_0$ and $c_1$ are scalars that account for the change in observable and units, and $\mathbf{\Delta}\in\mathbb{R}^{7}$ represents amplitude and phase offsets. The correction GP, $\delta^{\rm hist}(\x)$, used an RBF-7 kernel. The calibration and noise parameters, together with the RBF-7 kernel parameters, were fitted by minimizing the negative log likelihood, as in Eq.~\ref{eqn:neg_log_lik}, against a two-week historical validation dataset. During fitting, the simulation GP, $f(\x)$, remained fixed while the correction GP, $\delta^{\rm hist}(\x)$, was updated. We also examined a simulation-only case, for which we fitted the simulation GP to Eq. \eqref{eqn:sim_plus_historical}. This gave us a reliable connection between the simulation's percentage count and LANSCE's loss in its own units. This simulation calibration also included, for practical reasons, two points from a historical $\delta^{\rm hist}(\x)$, but these have negligible impact, since the simulation-only GP was trained on 2,000 points. We thus classify this method as Sim-Only-Cal.

During live tuning, we used the pretrained loss model $g(\x)$ to define the following latent GP for BO:
\begin{align}
 \underbrace{h(\x)}_{\text{full model}}
    &=
    \underbrace{g(\x)}_{\text{simulation/historic data}}
    +
    \underbrace{\delta^{\rm online }(\x)}_{\text{corrector}}.
\end{align}
The live loss observations satisfy:
\[
    L_t=h(\x_t)+\epsilon_t^{\rm online},
    \qquad
    \epsilon_t^{\rm online}\sim
    \mathcal{N}(0,\sigma_{\rm online}^{2}).
\]
We kept the pretrained model $g(\x)$ fixed while updating the correction GP, $\delta^{\rm online}(\x)$, as well as its parameters, and the measurement noise variance, $\sigma_{\rm online}^{2}$.

Three methods were evaluated using the fitting scheme in Eq. \eqref{eqn:sim_plus_historical}:
\begin{enumerate}
    \item \textbf{Sim-Only-Cal}: $\delta^{\rm{hist}}$ incorporated only two data points, whereas $f$ incorporated 2,000,
    \item \textbf{One-Week-Cal}: $\delta^{\rm{hist}}$ incorporated 1,000 points of historical data across a one-week period,
    \item \textbf{One-Month-Cal}: $\delta^{\rm{hist}}$ incorporated 1,000 points of historical data across a one-month period.
\end{enumerate}

We were unable to test a ``Three-Months-Cal'' method as the calibration (fitting as in Eq. \eqref{eqn:sim_plus_historical}) failed. This was an early indicator that older historical data (or at least large windows of historical data) perform poorly. 

Three other methods were constructed using an RBF-7 kernel to train a GP on 1,000 historical data points:
\begin{enumerate}
    \item \textbf{One-Week-Hist}: data from a week-long window,
    \item \textbf{One-Month-Hist}: data from a month-long window,
    \item \textbf{Three-Months-Hist}: data from a three-month-long window.
\end{enumerate}

During live testing, we used Badger's graphical user interface \cite{zhang2022badger} to connect the BO-ES-StepLmt algorithm and the GP models to the LANSCE accelerator, with all algorithms running on a CPU. The ES hyperparameters were unchanged from those used in simulation, except that the step sizes were adjusted to comply with safety requirements. Similarly, the ES updates were calculated by mapping the accelerator settings $x_j^i$ onto the interval $[-1,1]$, just like in simulation. This was done using the average and range of the global bounds, which were more stringent for the online test.
\subsection{Results and Analysis}

\begin{table*}[t]
\centering
\linespread{1}\selectfont 
\caption{DTL optimization statistics at LANSCE.
$L_t$ is the measured loss at iteration $t$, and $\bar{L}_0$
is the mean of the 42 initial measurements.
Uncertainties are propagated measurement uncertainties. $\langle \cdot \rangle_5$ represents a five-point window average. $\uparrow$ ($\downarrow$) indicates that higher (lower) values are better. }
\label{tab:LANSCE_results}

\setlength{\tabcolsep}{5pt}

\begin{tabular}{lccc}
\hline\hline
Method
& \shortstack{Avg. $L_t/\bar{L}_0$ $\downarrow$\\ (all)}
& \shortstack{Avg. $L_t/\bar{L}_0$ $\downarrow$\\ (only ES)}
& \shortstack{Avg. $L_t/\bar{L}_0$  $\downarrow$ \\ (last 20 iterations)} \\
\hline
Sim-Only-Cal
& $0.818 \pm 0.005$ & $0.797 \pm 0.005$
& $0.757 \pm 0.007$ \\
One-Week-Cal
& \best{0.808 \pm 0.005} & \best{0.769 \pm 0.005}
& $0.764 \pm 0.007$ \\
One-Week-Hist
& \best{0.802 \pm 0.005} & \best{0.769 \pm 0.005}
& \best{0.736 \pm 0.007} \\
One-Month-Cal
& $0.886 \pm 0.005$ & $0.827 \pm 0.005$
& \best{0.740 \pm 0.007} \\
One-Month-Hist
& $0.842 \pm 0.005$ & $0.789 \pm 0.005$
& $0.746 \pm 0.007$ \\
Three-Months-Hist
& $0.908 \pm 0.005$ & $0.856 \pm 0.005$
& $0.754 \pm 0.007$ \\
\hline\hline
\end{tabular}

\par\vspace{8pt}

\begin{tabular}{lccccc}
\hline\hline
Method
& \shortstack{Range $\langle L_t/\bar{L}_0\rangle_5$\\ (all)}
& \shortstack{Range $\langle L_t/\bar{L}_0\rangle_5$\\ (only BO)}
& \shortstack{Range $\langle L_t/\bar{L}_0\rangle_5$\\ (only ES)}
& \shortstack{$\max_t L_t/\bar{L}_0$ $\downarrow$\\ (only BO)} 
& \shortstack{$\max_t L_t/\bar{L}_0$ $\downarrow$\\ (last 100 iterations)} \\
\hline
Sim-Only-Cal
& $0.30 \pm 0.02$ & $0.19 \pm 0.02$ & $0.19 \pm 0.02$ & \best{1.05 \pm 0.04} & $0.89 \pm 0.03$ \\
One-Week-Cal
& $0.35 \pm 0.02$ & $0.18 \pm 0.02$ & $0.23 \pm 0.02$ & \best{1.06 \pm 0.04} & \best{0.85 \pm 0.03} \\
One-Week-Hist
& $0.37 \pm 0.02$ & $0.23 \pm 0.02$ & $0.21 \pm 0.02$ & \best{1.06\pm 0.04} & \best{0.82 \pm 0.03} \\
One-Month-Cal
& $0.46 \pm 0.02$ & $0.21 \pm 0.02$ & $0.40 \pm 0.02$ & $1.17 \pm 0.04$ & $0.93 \pm 0.03$ \\
One-Month-Hist
& $0.40 \pm 0.02$ & $0.23 \pm 0.02$ & $0.24 \pm 0.02$ & $1.13 \pm 0.04$ & \best{0.85 \pm 0.03} \\
Three-Months-Hist
& $0.54 \pm 0.02$ & $0.31 \pm 0.02$ & $0.54 \pm 0.02$ & $1.23 \pm 0.04$ & $0.86 \pm 0.03$ \\
\hline\hline
\end{tabular}
\end{table*}

\begin{figure*}
    \centering
    \includegraphics[width=\linewidth]{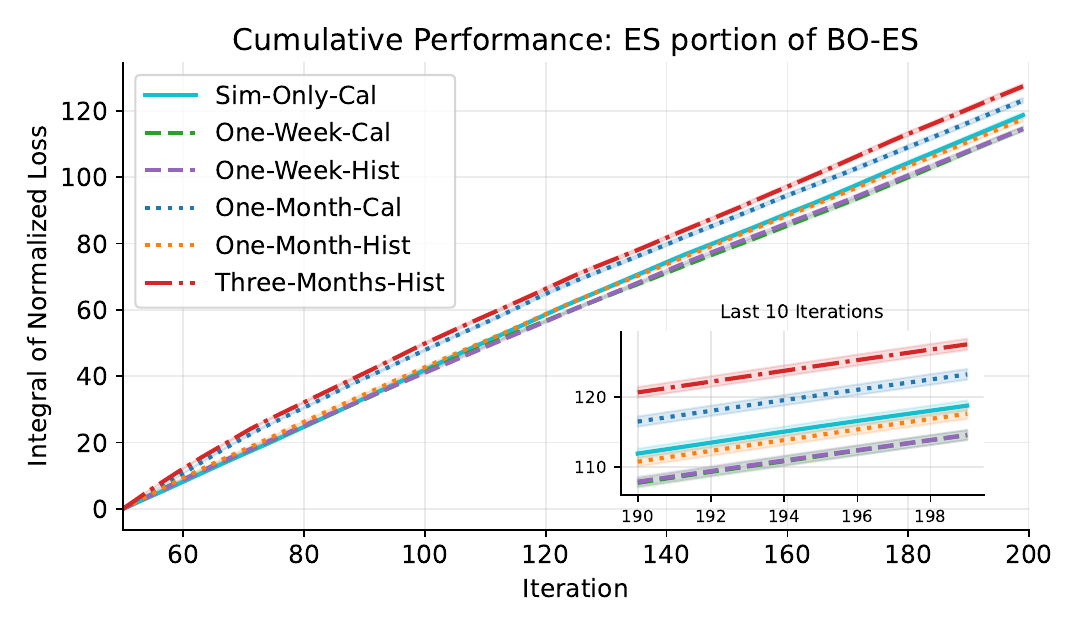}
    \caption{Integral of the normalized loss (normalized by $\bar L_0$) using different GPs for the BO. The color and line-style schemes mimic Fig. \ref{fig:Linac_Loss_BO_ES}. The algorithm changed from BO to ES at iteration 50. The lines have error bands for the integral uncertainty. The inset zooms into the accrued loss over the last ten iterations. We see that, as seen in Fig. \ref{fig:Linac_Loss_BO_ES}, the One-Month-Cal and Three-Months-Hist GPs produced substantially worse results. The other GPs led to better loss reductions over the ES phase, with the one-week GPs delivering the best results. }
    \label{fig:Linac_Cum_Loss_ES}
    \label{fig:cumulative_loss}
\end{figure*}

We began each run with seven iterations at identical amplitude and phase settings. This concentrated the initial data at a known safe operating point and provided each GP with repeated measurements to estimate noise within its run. Across the six runs, these 42 measurements also allowed us to estimate the measurement uncertainty under the assumption of homoscedastic noise. Unlike in simulation, noise-free values were unavailable. To estimate the underlying loss signal, we smoothed the loss time series using a centered moving average with a window of five observations.

Figure~\ref{fig:LANSCE_loss} shows the loss for each method, normalized by the initial loss, $\bar{L}_0$, estimated as the mean of the 42 initial measurements. Each GP was initialized using measurements at this known safe operating point. All methods reached final losses of approximately $70$--$80\%$ of the initial value. Incorporating more than one week of historical data generally led to poorer performance. The methods using only simulation data or one week of historical data generally reduced loss during the initial BO phase, whereas the other methods initially increased loss. This initial deterioration resembles the behavior observed for ES in simulation and in previous work using ES to tune the LANSCE DTL~\cite{scheinker2020extremum}, where optimization initially proceeded in the wrong direction (i.e., loss increased). However, the latter study used a more complex objective function based on readings from loss monitors at three locations.

These results suggest that the accelerator's drift timescale renders data older than one week outdated and less helpful for DTL tuning. This remains the case even if one uses the initial loss for each method rather than estimating it with the mean of the 42 initial measurements (see Appendix \ref{appendix:LANSCE_updates}). Assigning less weight to older observations may mitigate this effect. However, given BO's computational cost with large datasets, excluding data substantially older than one week and increasing the number of recent observations would likely be more computationally efficient. We limited each historical dataset to 1,000 observations to manage this cost. Future work could determine how much the sample size can be increased before the computational burden becomes prohibitive.

Using simulation data with calibration achieved performance comparable, albeit inferior, to that obtained by incorporating one week of historical data (Table~\ref{tab:LANSCE_results}). 
The coarseness of the historical amplitude data relative to the precision of the control suggests that the algorithms were not merely going to previous good set points. 

Figure~\ref{fig:cumulative_loss} shows the cumulative loss during the ES phase.

To determine whether the reduction in beam loss came at the expense of a significant reduction in transmitted beam current, we examined measurements from current monitors at the end of the linac. The transmitted currents remained largely stable, generally staying within approximately $2\%$ of their initial values (see Appendix~\ref{appendix:LANSCE_updates}).

BO-ES-StepLmt performed successfully despite the substantial modifications required for live accelerator operation, strongly suggesting that the method is robust. Switching to ES also enabled the One-Month-Cal and Three-Months-Hist runs to recover substantially from their poor initial performance, although their final losses remained higher than those of the other methods.

Most algorithm hyperparameters selected in simulation successfully transferred to live accelerator operation, demonstrating the value of the simulation trials. The principal exception was the step size limits, which required adjustment to comply with safety standards. We also increased the number of iterations from 150 to 200 just in case the methods took longer to converge in the more complicated case of the real world. Nonetheless, the algorithms operated smoothly. Valuable accelerator time was not wasted on hyperparameter tuning. 


\section{Conclusion}\label{section_label:Conclusion}

We have demonstrated the effectiveness of BO-ES for accelerator tuning in both simulation and live operation. Combining BO with ES allows historical and simulation data to be incorporated into the initial BO stage, providing information about the optimization landscape and guiding the algorithm toward a promising basin of solutions. ES can then take over, providing effective local optimization while handling drifting conditions very well. The imposed step limit in BO and the inherent step limit in bounded ES allow smooth parameter adjustments, as often required in accelerator operation. 

BO-ES can be extended in several ways. In this study, the BO search was locally constrained and myopic. Future work could investigate path-based and multi-step planning approaches that enable global exploration while still respecting step size limits \cite{folch2023practical, folch2024transition, paulson2022efficient}. Whether such methods can be useful in the drifting conditions of accelerators is an open question. For real-world usages of BO-ES, more attention would be needed for safety considerations, which were only addressed here via step size limits and global bounds. Further extensions could incorporate safety constraints by modifying both BO \cite{roussel2024bayesian} and ES \cite{williams2024experimental}. Performance constraints, like the desire for the transmitted current to be above a certain threshold, can also be included. 

Here we found that using historical-data windows longer than one week was detrimental. A more systematic study could also examine how best to combine simulation and historical data and select the time window over which the historical data are most useful. 

\begin{acknowledgments}
Los Alamos National Laboratory is operated by Triad National Security, LLC, for the national Nuclear Security Administration of the US Department of Energy (Contract No. 89233218CNA000001).
\end{acknowledgments}

\appendix
\section{Training Data and RBF Kernel Testing}\label{appendix:training_data_and_RBF}

\begin{figure*}
\centering
\includegraphics[width=\linewidth]{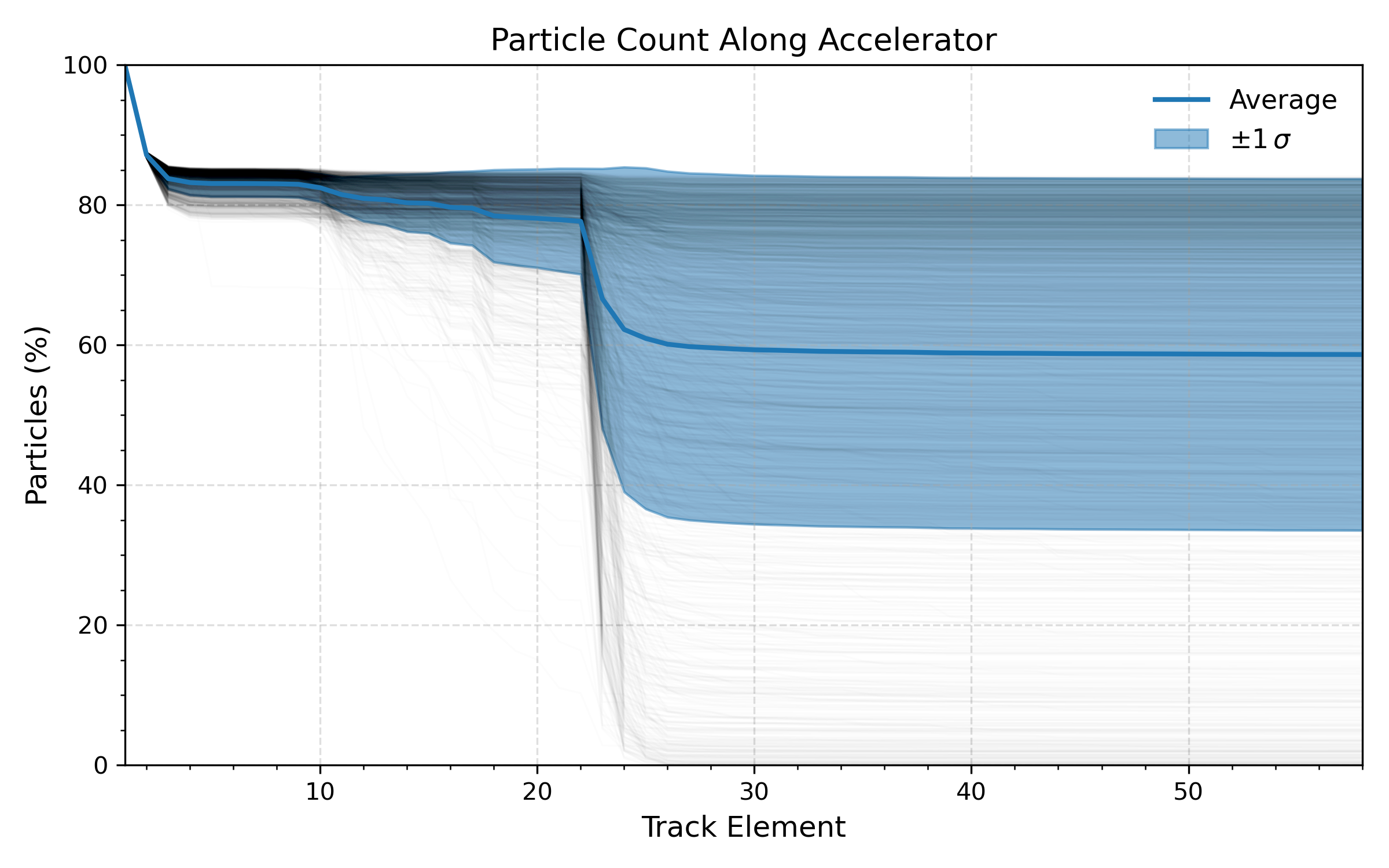}
\caption{ Simulation data for particle count along the linac section. The light gray curves represent different particle trajectories through the accelerator. The bright blue curve is the average of all trajectories and the shaded blue area shows the $1\sigma$ deviation away from it. Track elements correspond to DTL modules, CCL modules, and other apparatuses. }\label{fig:training_data}
\end{figure*}

\begin{figure*}
\centering
\includegraphics[width=0.9\linewidth]{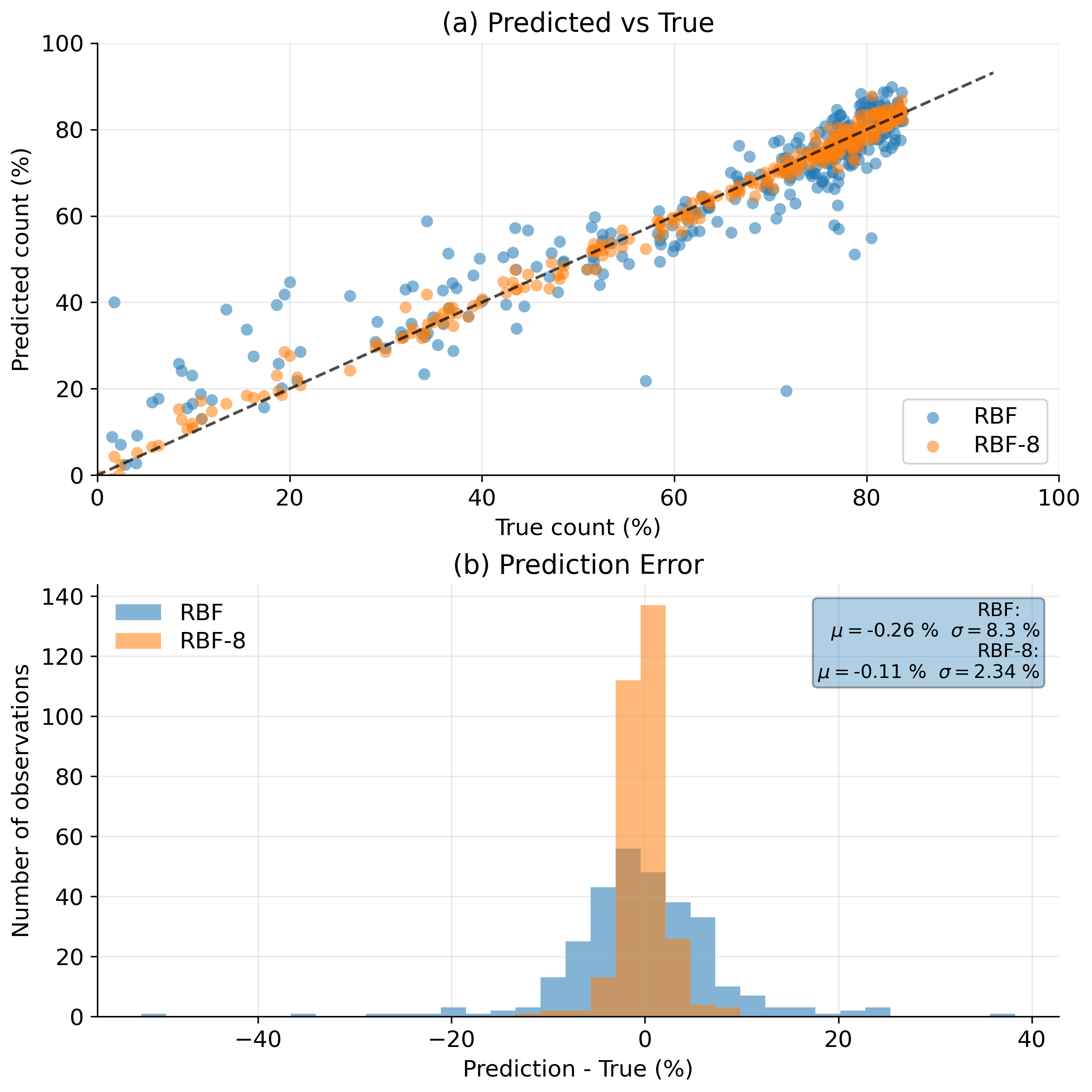}
\caption{Comparison of fits for different kernels. (a) A plot of true vs. predicted count for the two RBF kernels on the test set. (b) Histogram of errors for the two RBF kernels on the test set. }\label{fig:gp_fits_rbf1_v_rbf8}
\end{figure*}

The simulation dataset used for training GPs was made in  \texttt{HPSim}. Figure~\ref{fig:training_data} shows the particle count along the accelerator for many different runs with their own RF parameters. The particle count at the end was used to train the simulation surrogate GPs. For the kernel in Eq. \ref{eqn:rbf-kernel}, which we label RBF-8, each dimension gets its own dimension scaling, $\ell_i$. Instead, we can have a single scale for all dimensions:
\[
k(\x,\x') \;=\; \sigma_f^2 \exp\!\Big(-\tfrac{1}{2}\,\|\x-\x'\|^2/\ell^2\Big).
\]
We label this kernel RBF. We tested both kernels on the \texttt{HPSim} data. This used an older calibration. 2,000 data points were generated, uniformly sampled to be between $\pm3 \ \%$ (deg.) for the amplitude (phase) of the reference calibration. To test the fits, the GP models were trained on 85\% of the data and tested on the remaining 15\%. Figure \ref{fig:gp_fits_rbf1_v_rbf8} shows the results of the fit, with RBF-8 performing better. 

\begin{figure*}
\centering
\includegraphics[width=0.99\linewidth]{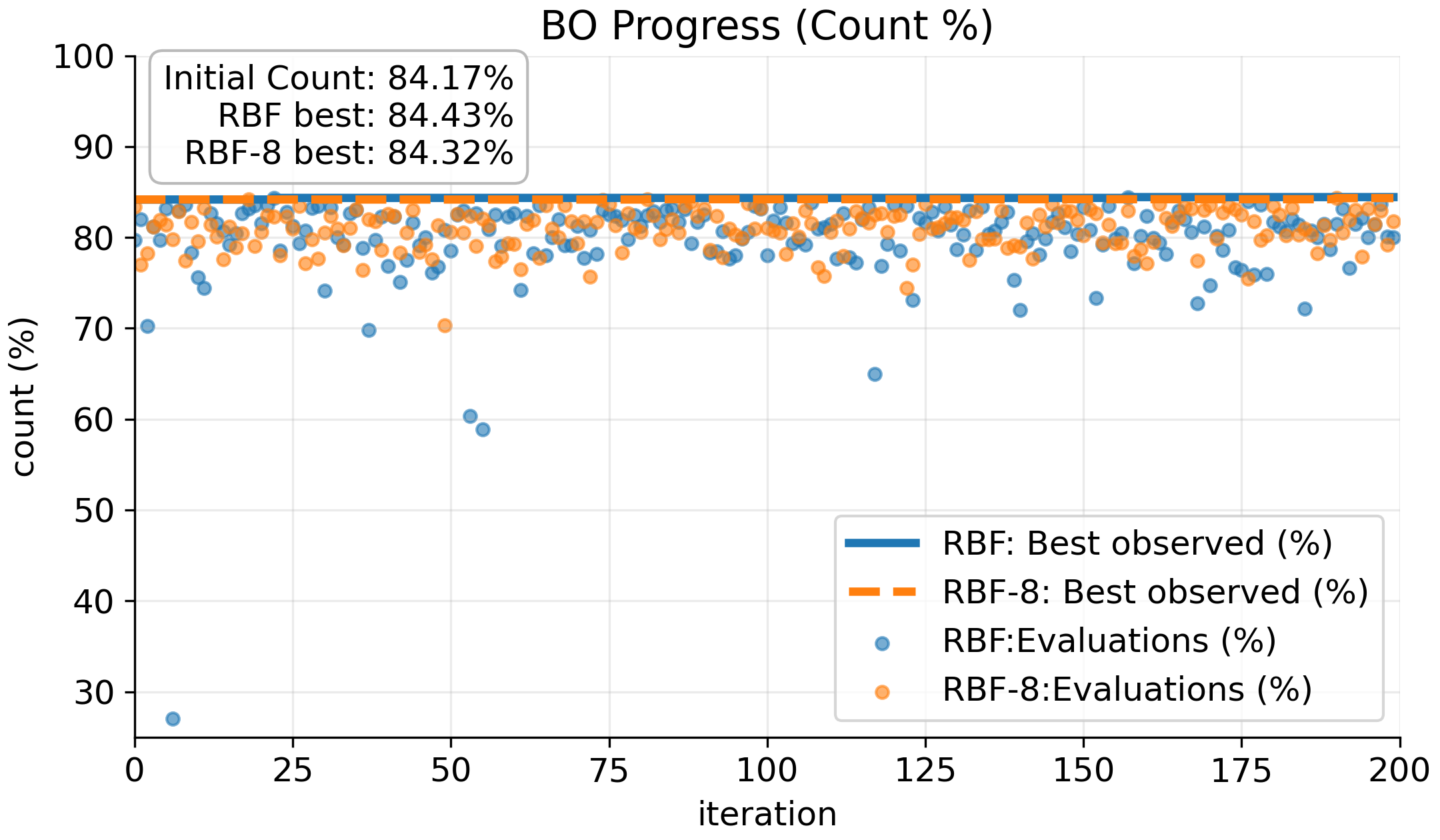}
\caption{A comparison of BO runs using GPs with RBF and RBF-8 kernels.}\label{fig:BO_on_different_RBF}
\end{figure*}

For one further test, BO was performed in simulation with both kernels, as seen in Fig. \ref{fig:BO_on_different_RBF}. Since the starting calibration point was already quite optimal, neither method increased the count by much. However, RBF produced much wilder swings, with the count falling below 30\%, which is far too low. RBF-8's low point was close to 70\%. Subsequently, we decided to go with RBF-8 for all experiments involving BO.

\section{Proof sketch for average gradient descent in 1D for ES}\label{appendix:proof_ES_grad_desc}

For a smooth, time-independent objective $J(x)$, the high-frequency averaged dynamics of Eq.~\ref{eqn:ES-varying} produce gradient descent on $J$.

\textit{Proof sketch for the 1D case}:

For compact notation, define the oscillation phase:
\begin{equation*}
    \theta(\tau)=\omega\tau+\kappa J(x(\tau)).
\end{equation*}
Subtract the high-frequency component by defining:
\begin{equation}\label{eqn:low_filter_x}
    X(\tau)=x(\tau)
    -\sqrt{\frac{2\alpha}{\omega}}\sin\theta(\tau).
\end{equation}
Then, from Eq.~\ref{eqn:ES-varying}:
\begin{align}
    \frac{dX}{d\tau}
    &=
    \frac{dx}{d\tau}
    -\sqrt{\frac{2\alpha}{\omega}}\cos\theta(\tau)
    \nonumber\\
    &\quad\times
    \left[
        \omega+\kappa J'(x(\tau))\frac{dx}{d\tau}
    \right]
    \nonumber\\
    &=
    -2\alpha\kappa J'(x(\tau))\cos^2\theta(\tau).
    \nonumber
\end{align}
Integrating:
\begin{align}
    X(\tau)
    &=
    X(0)-2\alpha\kappa
    \int_0^\tau
    J'(x(u))\cos^2\theta(u)\,du.
    \nonumber
\end{align}
Using $\cos^2\theta=(1+\cos2\theta)/2$:
\begin{align}\label{eqn:X_3_terms}
    X(\tau)
    &=
    X(0)-\alpha\kappa
    \int_0^\tau J'(x(u))\,du
    \nonumber\\
    &\quad-\alpha\kappa
    \int_0^\tau J'(x(u))\cos(2\theta(u))\,du.
\end{align}
In the high-frequency averaging approximation, the rapidly oscillating last term is neglected. This is motivated by the Riemann--Lebesgue lemma, which states that integrals of fixed integrable amplitudes multiplied by sinusoids tend to zero as the oscillation frequency tends to infinity, $\omega \rightarrow\infty$. Also, from Eq. \ref{eqn:low_filter_x},
\begin{equation*}
    |x(\tau)-X(\tau)|
    \leq\sqrt{\frac{2\alpha}{\omega}}.
\end{equation*}

To define the averaged trajectory precisely, first average the drift over one oscillation cycle while holding the slow coordinate $z$ fixed:
\begin{align}
    \bar F(z)
    &=
    -\frac{2\alpha\kappa J'(z)}{2\pi}
    \int_0^{2\pi}\cos^2\vartheta\,d\vartheta
    \nonumber\\
    &=-\alpha\kappa J'(z).
    \nonumber
\end{align}
We define $\bar X(\tau)$ as the solution of the averaged initial-value problem:
\begin{equation}\label{eqn:X_grad_desc}
        \frac{d\bar X}{d\tau}
        =\bar F(\bar X)
        =-\alpha\kappa J'(\bar X).
\end{equation}
Its initial condition is $\bar X(0)=x(0)$, since the difference between $X(0)$ and $x(0)$ vanishes as $\omega\to\infty$.

Thus, $\bar X$ describes the slow motion after averaging over the rapid oscillations, and its dynamics are gradient descent on $J$. 

\section{Noisy Measurement Plots}\label{appendix:noisy_measurement}

\begin{figure*}
\centering
\includegraphics[width=\linewidth]{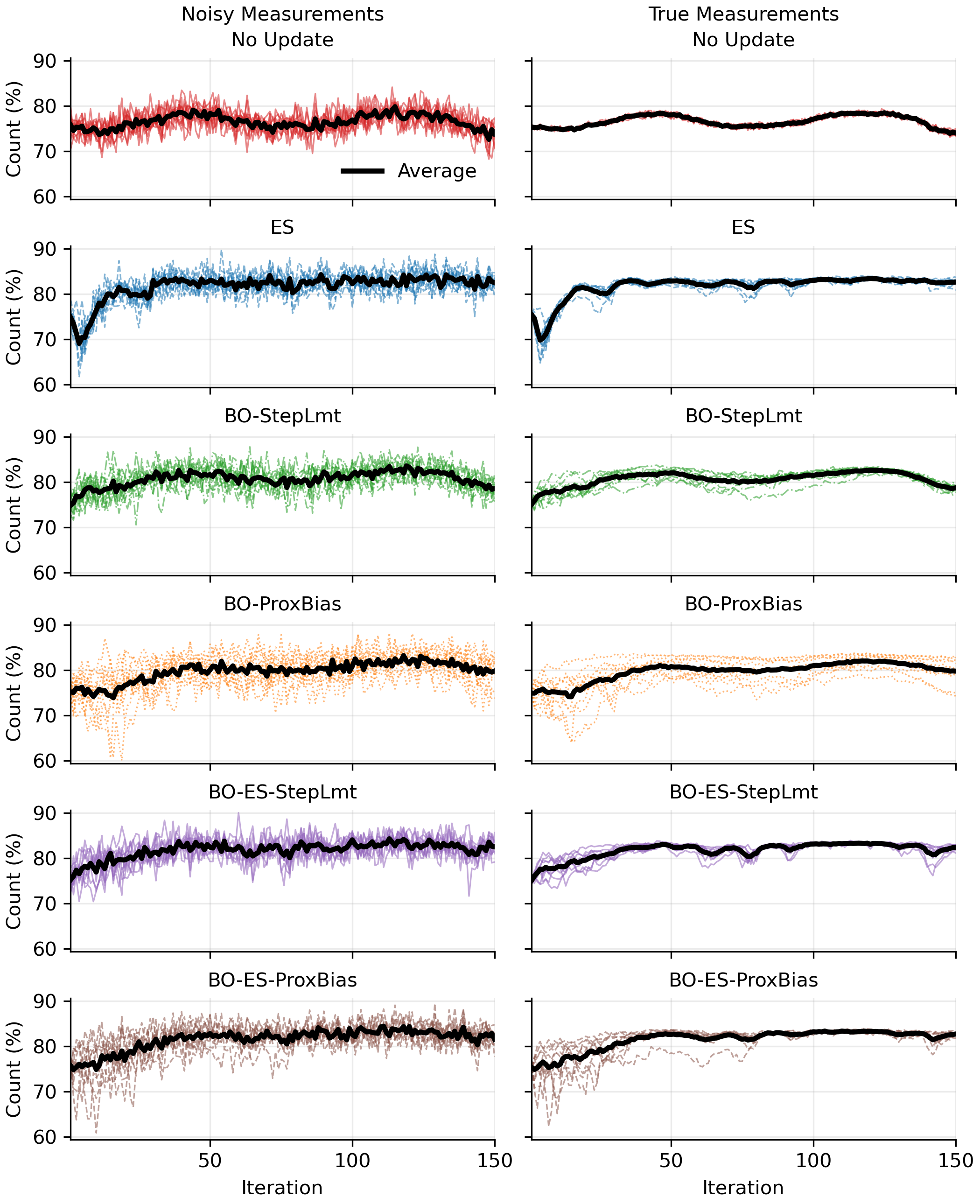}
\caption{BO progression under sinusoidal drift conditions. The left plots show the noisy measurements the algorithms see, while the right plots show the true measurement values. All ten runs for each method are shown.}\label{fig:drift_noisy_progress}
\end{figure*}

Figure \ref{fig:drift_noisy_progress} shows the noisy vs. true measurements for each method in the sinusoidal drift case. 

\section{Amplitudes, phases, current and adjusted loss during LANSCE DTL Tuning}\label{appendix:LANSCE_updates}

\begin{figure*}
    \centering
    \includegraphics[width=0.48\linewidth]{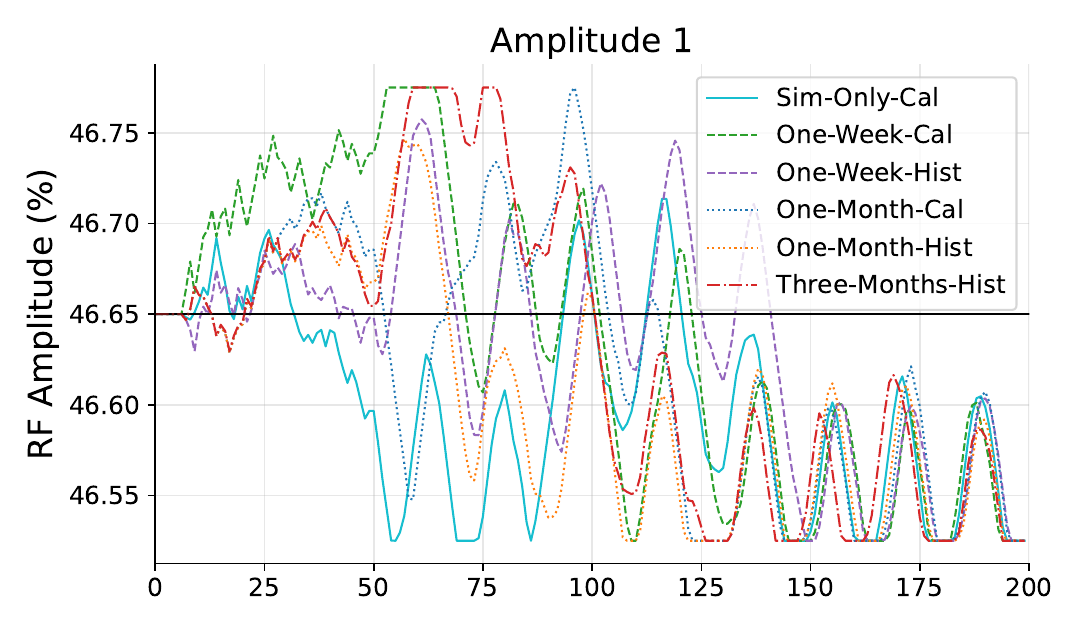}%
    \includegraphics[width=0.48\linewidth]{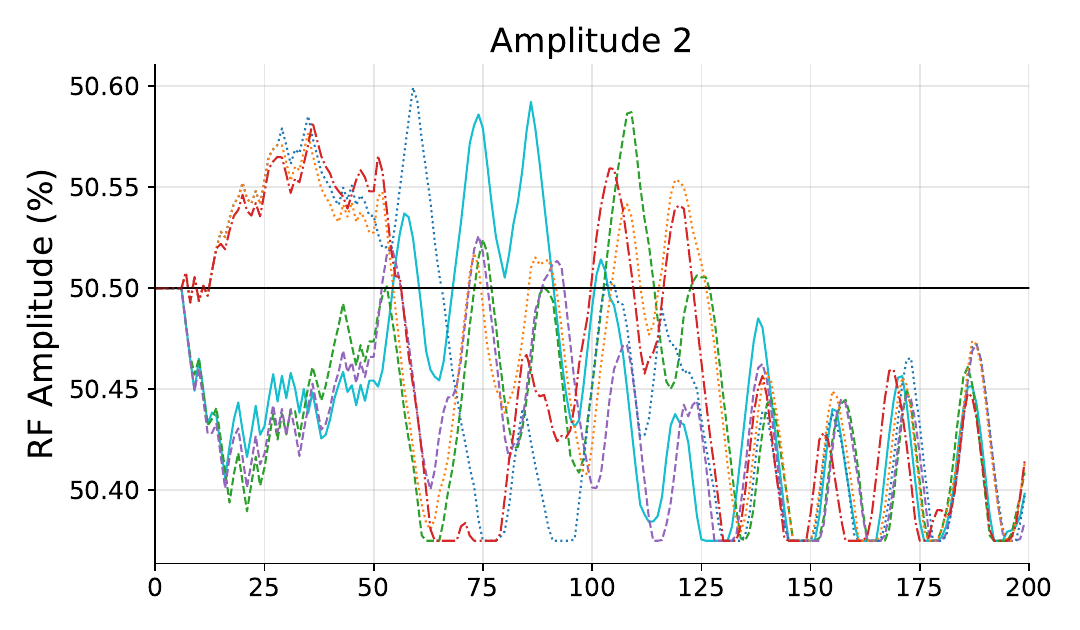}
    \includegraphics[width=0.48\linewidth]{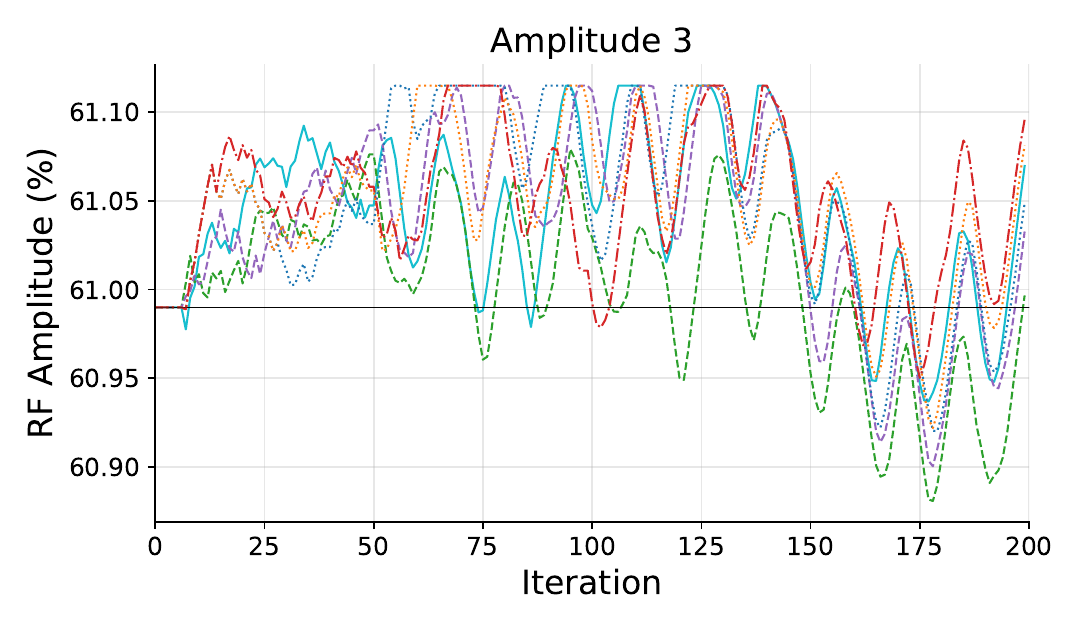}%
    \includegraphics[width=0.48\linewidth]{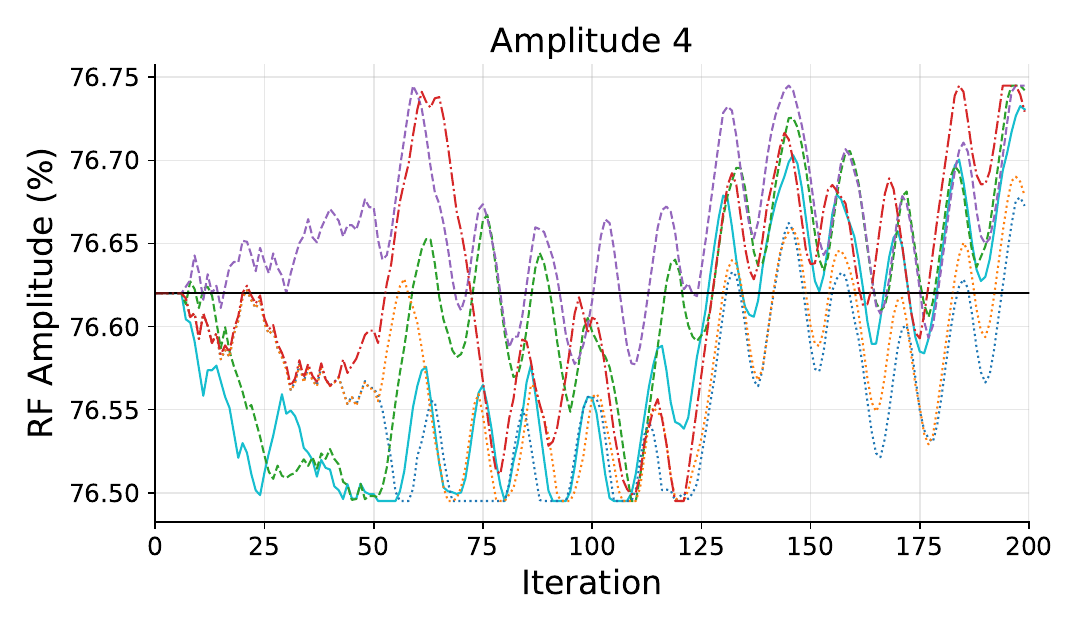}
    \caption{RF amplitudes suggested by the BO-ES algorithm at each iteration for each of the four DTL modules. Labels follow the scheme from Fig. \ref{fig:Linac_Loss_BO_ES}. Note that all amplitudes were held fixed for the first seven iterations; the horizontal black line represents this initial value in each plot. The transition to ES at the $50^{\rm th}$ iteration is visible by the change to a sinusoidal sweep of the values. }
    \label{fig:DTL_amplitudes}
\end{figure*}

\begin{figure*}
    \centering
    \includegraphics[width=0.48\linewidth]{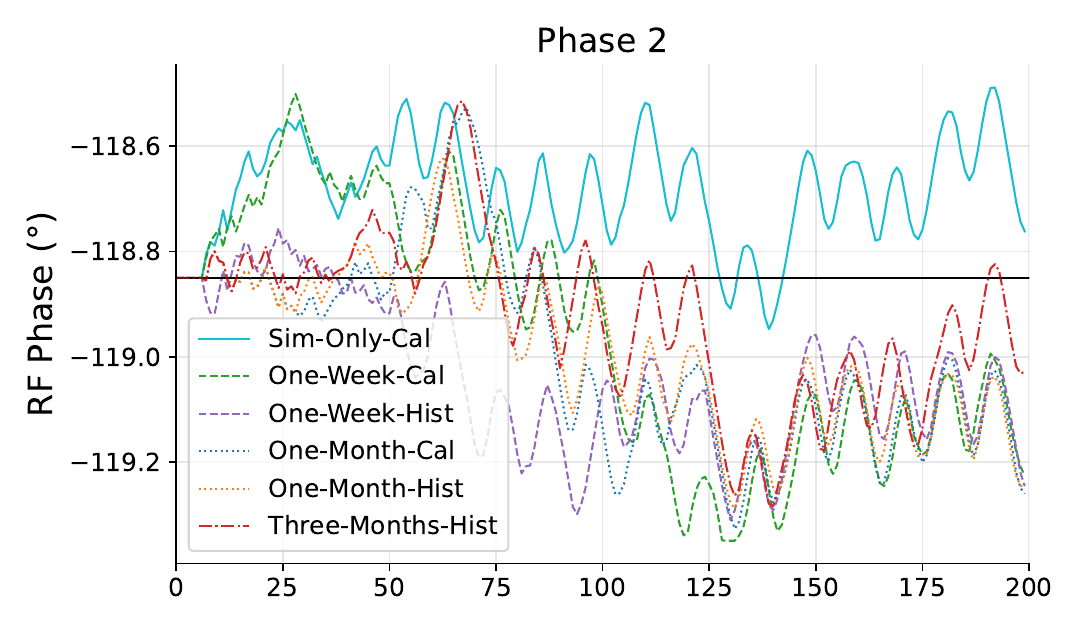}%
    \includegraphics[width=0.48\linewidth]{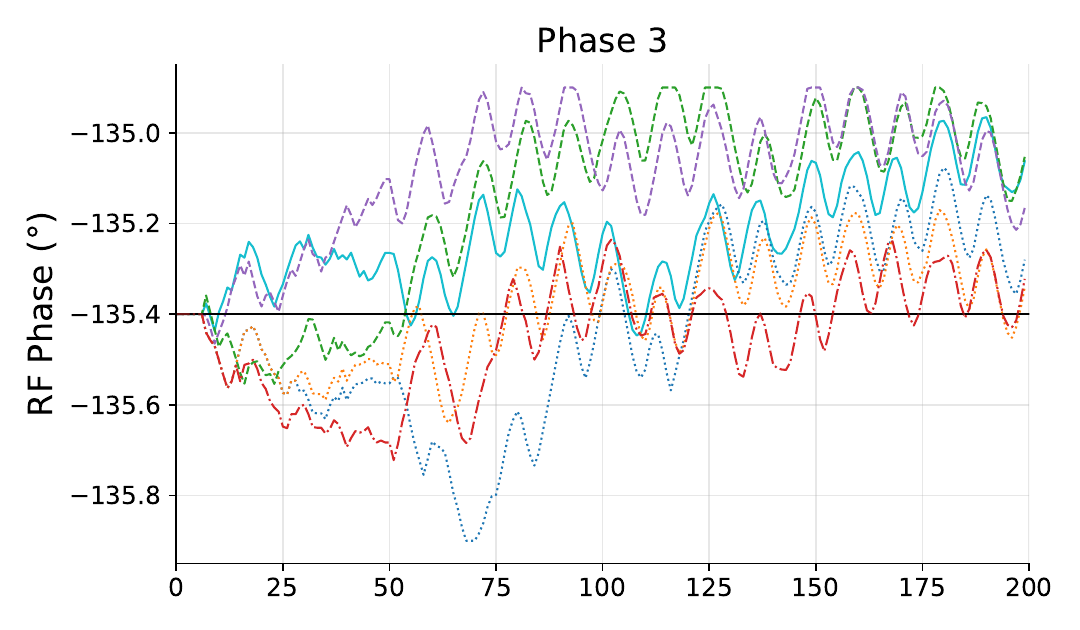}
    \includegraphics[width=0.48\linewidth]{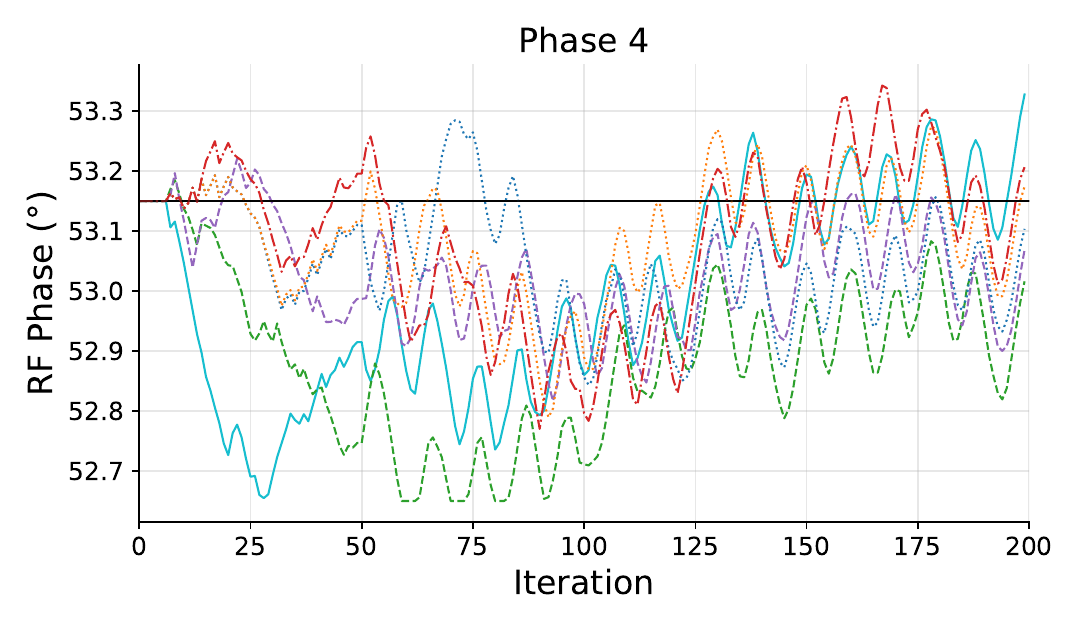}%
    \includegraphics[width=0.48\linewidth]{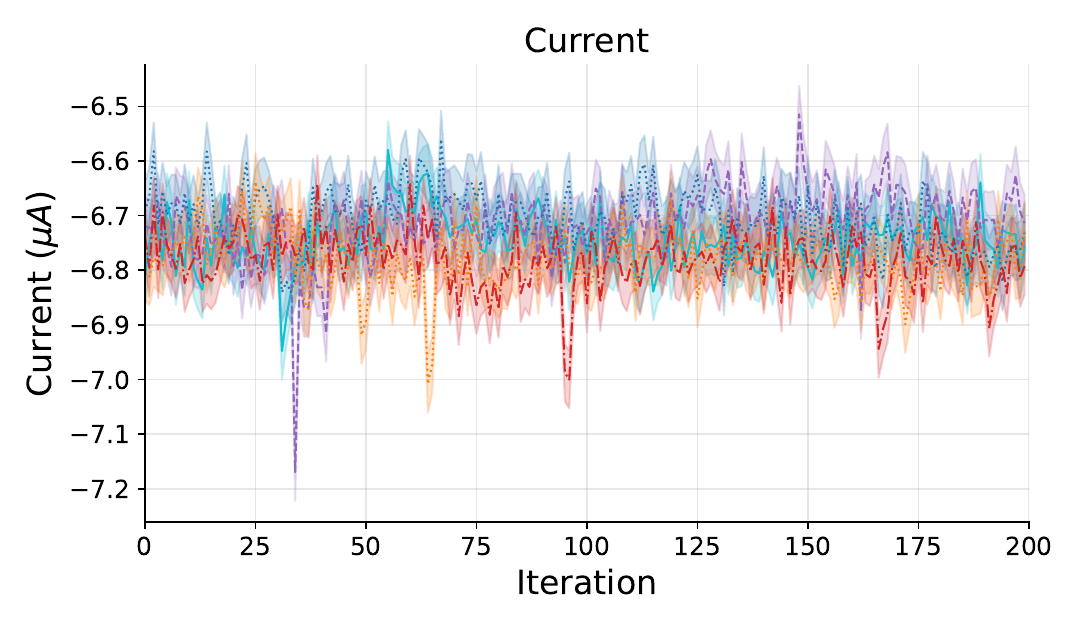}
    \caption{RF phases suggested by the BO-ES algorithm at each iteration for each of the last three DTL modules. The bottom right graph shows the measured current at the end of the 44 CCL modules, along with a 1$\sigma$ uncertainty band. Labels follow the scheme from Fig. \ref{fig:Linac_Loss_BO_ES}. Note that all phases were held fixed for the first seven iterations; the horizontal black line represents this initial value in each phase plot. The transition to ES at the $50^{\rm th}$ iteration is visible by the change to a sinusoidal sweep of the values. We removed a current outlier at iteration 40 of the One-Week-Hist run due to a momentary malfunction of one of the CCL modules downstream from our experiment. There is no current reading for the One-Week-Cal run because we did not record it.}
    \label{fig:DTL_phases_current}
\end{figure*}

The change in the amplitudes for each method tried is depicted in Fig. \ref{fig:DTL_amplitudes}. The phases and current are shown in Fig. \ref{fig:DTL_phases_current}. It is seen that the current remains steady across iterations both between and 
within each method.

\begin{figure*}
    \centering
    \includegraphics[width=\linewidth]{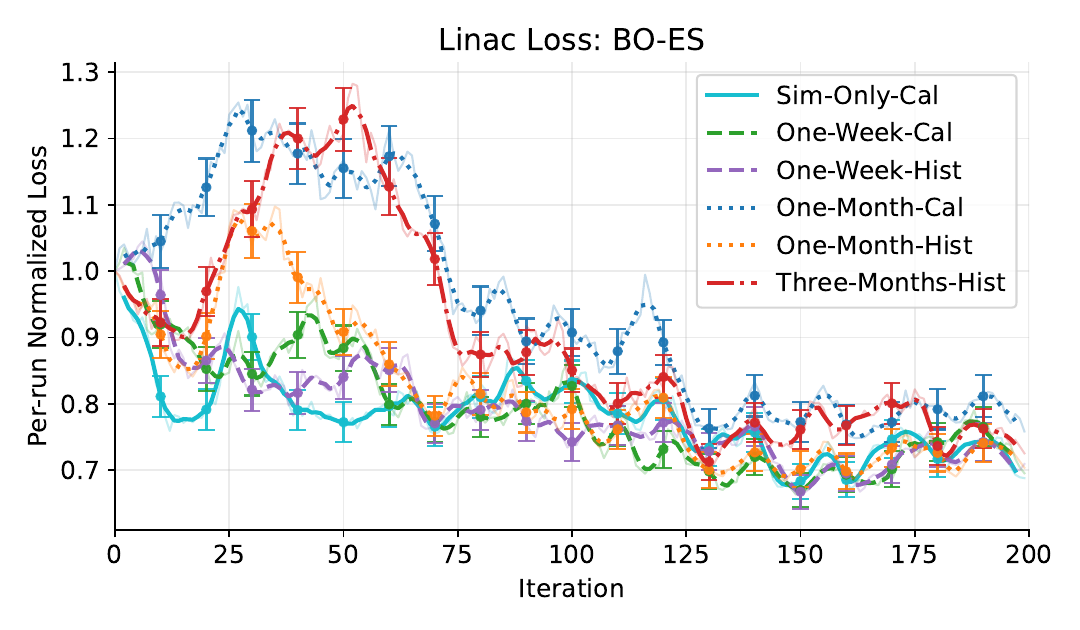}
    \caption{Five-point window average of the per-run normalized loss. Faint lines in the backdrop show the raw value for each iteration. Error bars represent $1\sigma$ of uncertainty for the window average and are only displayed every $10^{\rm th}$ iteration for visual clarity. Here, we normalized the loss by the first data point of that specific run. This is why the raw data all start at 1. In comparison to Fig. \ref{fig:Linac_Loss_BO_ES}, Sim-Only-Cal performs slightly better than the one-week GPs. However, the general conclusions about the robustness of BO-ES and of Sim-Only-Cal, One-Week-Hist, and One-Week-Cal being the best-performing GPs remain valid. Errors are larger than those in Fig. \ref{fig:Linac_Loss_BO_ES} due to increased uncertainty in the normalization constant. That is, normalization by 42 data points is more robust than normalization by an initial value.}
    \label{fig:per-run_linac_loss_BO_ES}
\end{figure*}

Figure \ref{fig:per-run_linac_loss_BO_ES} shows the normalized loss if one normalizes by the initial loss per run rather than normalizing by an estimate of the 42 initial measurements from the 6 runs. The performance at the beginning is still better for the Sim-Only-Cal and the one-week cases.

\bibliography{sample}

@article{hwang2211prior,
  title={Prior-mean-assisted bayesian optimization application on frib front-end tunning (2022)},
  author={Hwang, Kilean and Maruta, Tomofumi and Plastun, Alexander and Fukushima, Kei and Zhang, Tong and Zhao, Qiang and Ostroumov, Peter and Hao, Yue},
  journal={arXiv preprint arXiv:2211.06400}
}

@article{hwang2026customized,
  title={Customized Bayesian optimization for efficient beam tuning at the facility for rare isotope beams},
  author={Hwang, Kilean and Zhao, Qiang and Zhang, Tong and Maruta, Tomofumi and Plastun, Alexander and Fukushima, Kei and Ostroumov, Peter},
  journal={Physical Review Accelerators and Beams},
  volume={29},
  number={2},
  pages={024601},
  year={2026},
  publisher={APS}
}

@article{scheinker2016bounded,
  title={Bounded extremum seeking with discontinuous dithers},
  author={Scheinker, Alexander and Scheinker, David},
  journal={Automatica},
  volume={69},
  pages={250--257},
  year={2016},
  publisher={Elsevier}
}

@article{scheinker2019model,
  title={Model-independent tuning for maximizing free electron laser pulse energy},
  author={Scheinker, Alexander and Bohler, Dorian and Tomin, Sergey and Kammering, Raimund and Zagorodnov, Igor and Schlarb, Holger and Scholz, Matthias and Beutner, Bolko and Decking, Winfried},
  journal={Physical review accelerators and beams},
  volume={22},
  number={8},
  pages={082802},
  year={2019},
  publisher={APS}
}

@article{duris2020bayesian,
  title={Bayesian optimization of a free-electron laser},
  author={Duris, Joseph and Kennedy, Dylan and Hanuka, Adi and Shtalenkova, Jane and Edelen, Auralee and Baxevanis, P and Egger, Adam and Cope, T and McIntire, M and Ermon, S and others},
  journal={Physical review letters},
  volume={124},
  number={12},
  pages={124801},
  year={2020},
  publisher={APS}
}

@article{pang2014gpu,
  title={GPU accelerated online multi-particle beam dynamics simulator for ion linear particle accelerators},
  author={Pang, Xiaoying and Rybarcyk, Larry},
  journal={Computer Physics Communications},
  volume={185},
  number={3},
  pages={744--753},
  year={2014},
  publisher={Elsevier}
}

@inproceedings{roussel2021proximal,
  title={Proximal biasing for Bayesian optimization and characterization of physical systems},
  author={Roussel, Ryan and Edelen, Auralee},
  booktitle={Workshop on Machine Learning and the Physical Sciences},
  year={2021}
}

@article{kirschner2022tuning,
  title={Tuning particle accelerators with safety constraints using Bayesian optimization},
  author={Kirschner, Johannes and Mutn{\`y}, Mojmir and Krause, Andreas and Coello de Portugal, Jaime and Hiller, Nicole and Snuverink, Jochem},
  journal={Physical Review Accelerators and Beams},
  volume={25},
  number={6},
  pages={062802},
  year={2022},
  publisher={APS}
}

@article{roussel2024bayesian,
  title={Bayesian optimization algorithms for accelerator physics},
  author={Roussel, Ryan and Edelen, Auralee L and Boltz, Tobias and Kennedy, Dylan and Zhang, Zhe and Ji, Fuhao and Huang, Xiaobiao and Ratner, Daniel and Garcia, Andrea Santamaria and Xu, Chenran and others},
  journal={Physical Review Accelerators and Beams},
  volume={27},
  number={8},
  pages={084801},
  year={2024},
  publisher={APS}
}

@article{jalas2021bayesian,
  title={Bayesian optimization of a laser-plasma accelerator},
  author={Jalas, S{\"o}ren and Kirchen, Manuel and Messner, Philipp and Winkler, Paul and H{\"u}bner, Lars and Dirkwinkel, Julian and Schnepp, Matthias and Lehe, Remi and Maier, Andreas R},
  journal={Physical Review Letters},
  volume={126},
  number={10},
  pages={104801},
  year={2021},
  publisher={APS}
}

@article{boltz2025leveraging,
  title={Leveraging prior mean models for faster Bayesian optimization of particle accelerators},
  author={Boltz, Tobias and Martinez, Jose L and Xu, Connie and Baker, Kathryn RL and Zhu, Zihan and Morgan, Jenny and Roussel, Ryan and Ratner, Daniel and Mustapha, Brahim and Edelen, Auralee L},
  journal={Scientific Reports},
  volume={15},
  number={1},
  pages={12232},
  year={2025},
  publisher={Nature Publishing Group UK London}
}

@article{williams2024experimental,
  title={Experimental safe extremum seeking for accelerators},
  author={Williams, Alan and Scheinker, Alexander and Huang, En-Chuan and Taylor, Charles and Krstic, Miroslav},
  journal={IEEE Transactions on Control Systems Technology},
  volume={32},
  number={5},
  pages={1881--1890},
  year={2024},
  publisher={IEEE}
}

@article{thompson1933likelihood,
  title={On the likelihood that one unknown probability exceeds another in view of the evidence of two samples},
  author={Thompson, William R},
  journal={Biometrika},
  volume={25},
  number={3/4},
  pages={285--294},
  year={1933},
  publisher={JSTOR}
}

@article{thompson1935theory,
  title={On the theory of apportionment},
  author={Thompson, William R},
  journal={American Journal of Mathematics},
  volume={57},
  number={2},
  pages={450--456},
  year={1935},
  publisher={JSTOR}
}

@inproceedings{kandasamy2018parallelised,
  title={Parallelised Bayesian optimisation via Thompson sampling},
  author={Kandasamy, Kirthevasan and Krishnamurthy, Akshay and Schneider, Jeff and P{\'o}czos, Barnab{\'a}s},
  booktitle={International conference on artificial intelligence and statistics},
  pages={133--142},
  year={2018},
  organization={PMLR}
}

@inproceedings{scheinker2020extremum,
  title={Extremum Seeking for Minimization of Beam Loss in the LANSCE Linear Accelerator by Tuning RF Cavities},
  author={Scheinker, Alexander and Naffziger, Peter and Garcia, Antonio},
  booktitle={2020 American Control Conference (ACC)},
  pages={5071--5075},
  year={2020},
  organization={IEEE}
}

@book{scheinker2017model,
  title={Model-free stabilization by extremum seeking},
  author={Scheinker, Alexander and Krsti{\'c}, Miroslav},
  year={2017},
  publisher={Springer}
}

@article{chakrabarty2021accelerating,
  title={Accelerating self-optimization control of refrigerant cycles with Bayesian optimization and adaptive moment estimation},
  author={Chakrabarty, Ankush and Danielson, Claus and Bortoff, Scott A and Laughman, Christopher R},
  journal={Applied Thermal Engineering},
  volume={197},
  pages={117335},
  year={2021},
  publisher={Elsevier}
}

@misc{botorch,
  author       = {Balandat, Maximilian and Karrer, Brian and Jiang, Daniel and Daulton, Samuel and Letham, Benjamin and Wilson, Andrew Gordon and Bakshy, Eytan},
  title        = {BoTorch: A Framework for Efficient Monte-Carlo Bayesian Optimization},
  year         = {2020},
  eprint       = {1910.06403},
  archivePrefix= {arXiv},
  primaryClass = {cs.LG},
  url          = {https://arxiv.org/abs/1910.06403}
}

@inproceedings{gardner2018gpytorch,
  title     = {GPyTorch: Blackbox Matrix-Matrix Gaussian Process Inference with GPU Acceleration},
  author    = {Gardner, Jacob R. and Pleiss, Geoff and Weinberger, Kilian Q. and Bindel, David and Wilson, Andrew Gordon},
  booktitle = {Advances in Neural Information Processing Systems},
  year      = {2018}
}

@article{2020SciPy-NMeth,
  author  = {Virtanen, Pauli and others},
  title   = {{SciPy} 1.0: Fundamental Algorithms for Scientific Computing in Python},
  journal = {Nature Methods},
  volume  = {17},
  pages   = {261--272},
  year    = {2020},
  doi     = {10.1038/s41592-019-0686-2}
}

@article{10.1115/1.3653121,
    author = {Kushner, H. J.},
    title = {A New Method of Locating the Maximum Point of an Arbitrary Multipeak Curve in the Presence of Noise},
    journal = {Journal of Basic Engineering},
    volume = {86},
    number = {1},
    pages = {97-106},
    year = {1964},
    month = {03},
    issn = {0021-9223},
    doi = {10.1115/1.3653121},
    url = {https://doi.org/10.1115/1.3653121}
}

@InProceedings{10.1007/3-540-07165-2_55,
author="Mo{\v{c}}kus, J.",
editor="Marchuk, G. I.",
title="On bayesian methods for seeking the extremum",
booktitle="Optimization Techniques IFIP Technical Conference Novosibirsk, July 1--7, 1974",
year="1975",
publisher="Springer Berlin Heidelberg",
address="Berlin, Heidelberg",
pages="400--404",
isbn="978-3-540-37497-8"
}

@article{jones1998efficient,
  title={Efficient global optimization of expensive black-box functions},
  author={Jones, Donald R and Schonlau, Matthias and Welch, William J},
  journal={Journal of Global optimization},
  volume={13},
  number={4},
  pages={455--492},
  year={1998},
  publisher={Springer}
}

@incollection{rasmussen2003gaussian,
  title={Gaussian processes in machine learning},
  author={Rasmussen, Carl Edward},
  booktitle={Summer school on machine learning},
  pages={63--71},
  year={2003},
  publisher={Springer}
}

@article{engel1968geometrical,
  title={Geometrical and Electromagnetic Parameters of the Accelerating and Coupling Cells of the 805 MHz Linac for LAMPF},
  author={Engel, LN},
  journal={LANL, Los Alamos, NM, USA, Rep. MP-3-58},
  year={1968}
}

@article{scheinker2018demonstration,
  title={Demonstration of model-independent control of the longitudinal phase space of electron beams in the linac-coherent light source with femtosecond resolution},
  author={Scheinker, Alexander and Edelen, Auralee and Bohler, Dorian and Emma, Claudio and Lutman, Alberto},
  journal={Physical Review Letters},
  volume={121},
  number={4},
  pages={044801},
  year={2018},
  publisher={APS}
}

@article{scheinker2021extremum,
  title={Extremum seeking-based control system for particle accelerator beam loss minimization},
  author={Scheinker, Alexander and Huang, En-Chuan and Taylor, Charles},
  journal={IEEE Transactions on Control Systems Technology},
  volume={30},
  number={5},
  pages={2261--2268},
  year={2021},
  publisher={IEEE}
}

@article{scheinker2021adaptive,
  title={An adaptive approach to machine learning for compact particle accelerators},
  author={Scheinker, Alexander and Cropp, Frederick and Paiagua, Sergio and Filippetto, Daniele},
  journal={Scientific Reports},
  volume={11},
  number={1},
  pages={19187},
  year={2021},
  publisher={Nature Publishing Group UK London}
}

@article{roussel2023xopt,
  title={Xopt: A simplified framework for optimization of accelerator problems using advanced algorithms},
  author={Roussel, Ryan and Edelen, A and Bartnik, A and Mayes, Christopher},
  journal={Proc. IPAC},
  volume={23},
  number={05},
  pages={4796--4799},
  year={2023}
}

@misc{BatyginUSPAS2019_AccStruct,
  author = {Batygin, Yuri},
  title  = {Introduction to Accelerating Structures},
  year   = {2019},
  month  = jun,
  note   = {Lecture notes for Proton and Ion Linear Accelerators, U.S. Particle Accelerator School (USPAS), Albuquerque, New Mexico, June 17--28, 2019. LA-UR-16-29625. Accessed 2026-02-20},
  url    = {https://uspas.fnal.gov/materials/19NewMexico/PILA/2.Introduction_to_Accelerating_Structures.pdf}
}

@article{morita2023accelerator,
  title={Accelerator tuning method using autoencoder and Bayesian optimization},
  author={Morita, Yasuyuki and Washio, Takashi and Nakashima, Yuta},
  journal={Nuclear Instruments and Methods in Physics Research Section A: Accelerators, Spectrometers, Detectors and Associated Equipment},
  volume={1057},
  pages={168730},
  year={2023},
  publisher={Elsevier}
}

@article{zhang2022badger,
  title={Badger: The missing optimizer in ACR},
  author={Zhang, Zhe and Edelen, Auralee and Mayes, C and Garrahan, J and Shtalenkova, J and Roussel, R and Miskovich, S and Ratner, Daniel and Boese, Michael and Tomin, Sergey and others},
  journal={Proc. IPAC’22},
  pages={999--1002},
  year={2022},
  publisher={JACoW Publishing Geneva, Switzerland}
}

@inproceedings{paulson2022efficient,
  title={Efficient multi-step lookahead Bayesian optimization with local search constraints},
  author={Paulson, Joel A and Sorouifar, Farshud and Chakrabarty, Ankush},
  booktitle={2022 IEEE 61st Conference on Decision and Control (CDC)},
  pages={123--129},
  year={2022},
  organization={IEEE}
}

@article{folch2023practical,
  title={Practical path-based Bayesian optimization},
  author={Folch, Jose Pablo and Odgers, James and Zhang, Shiqiang and Lee, Robert M and Shafei, Behrang and Walz, David and Tsay, Calvin and van der Wilk, Mark and Misener, Ruth},
  journal={arXiv preprint arXiv:2312.00622},
  year={2023}
}

@article{folch2024transition,
  title={Transition constrained Bayesian optimization via Markov decision processes},
  author={Folch, Jose P and Tsay, Calvin and Lee, Robert M and Shafei, Behrang and Ormaniec, Weronika and Krause, Andreas and van der Wilk, Mark and Misener, Ruth and Mutn{\`y}, Mojm{\'\i}r},
  journal={Advances in Neural Information Processing Systems},
  volume={37},
  pages={88194--88235},
  year={2024}
}

@article{eriksson2019scalable,
  title={Scalable global optimization via local Bayesian optimization},
  author={Eriksson, David and Pearce, Michael and Gardner, Jacob and Turner, Ryan D and Poloczek, Matthias},
  journal={Advances in neural information processing systems},
  volume={32},
  year={2019}
}

@article{scheinker2026adaptive,
  title={Adaptive control meets deep learning: Extremum seeking in latent space with particle accelerator applications},
  author={Scheinker, Alexander},
  journal={IEEE Control Systems},
  volume={46},
  number={2},
  pages={150--176},
  year={2026},
  publisher={IEEE}
}

@inproceedings{saxena2026improved,
  title={Improved robustness of deep reinforcement learning for control of time-varying systems by bounded extremum seeking},
  author={Saxena, Shaifalee and Williams, Alan and Fierro, Rafael and Scheinker, Alexander},
  booktitle={2026 American Control Conference (ACC)},
  pages={1165--1172},
  year={2026},
  organization={IEEE}
}

@article{saxena2026mahalanobis,
  title={Mahalanobis-Guided Latent OOD Detection for Hybrid ES-DRL Control in Time-Varying Systems},
  author={Saxena, Shaifalee and Scheinker, Alexander},
  journal={arXiv preprint arXiv:2606.11474},
  year={2026}
}

@article{mahalanobis2018generalized,
  title={On the generalized distance in statistics},
  author={Mahalanobis, Prasanta Chandra},
  journal={Sankhy{\=a}: The Indian Journal of Statistics, Series A (2008-)},
  volume={80},
  pages={S1--S7},
  year={2018},
  publisher={JSTOR}
}

@article{sobol1967distribution,
  title={Distribution of points in a cube and approximate evaluation of integrals},
  author={Sobol, Ilya M},
  journal={USSR Computational mathematics and mathematical physics},
  volume={7},
  pages={86--112},
  year={1967}
}

\end{document}